\documentclass[12pt, preprint]{aastex}
\usepackage{amsmath,natbib}
\received{2004 December 2}
\revised{2005 July 12}
\accepted{2005 July 12}
\begin{document}

\title{Neptune's Migration into a Stirred--Up Kuiper Belt:\\
A Detailed Comparison of Simulations to Observations\vspace*{0.05in}}

\author{
Joseph M. Hahn\altaffilmark{1}
\altaffiltext{1}{Canada Research Chair in Astrophysics}
}
\affil{
  Institute for Computational Astrophysics\\
  Department of Astronomy and Physics\\
  Saint Mary's University\\
  Halifax, NS, B3H 3C3, Canada\\
  email: jhahn@ap.stmarys.ca\\
  phone: 902--420--5428\\
  fax  : 902--420--5141\vspace*{0.15in}
}

\author{Renu Malhotra}
\affil{
  University of Arizona\\
  Department of Planetary Sciences\\
  1629 East University Boulevard\\
  Tuscon, AZ 85721--0092\\
  email: renu@lpl.arizona.edu\\
  phone: 520--626--5899\\
  fax:   520--626--8250\vspace{0.15in}
}

\author{
  Submitted to the {\it Astronomical Journal}, 2004 December 2\\
  Accepted 2005 July 12
}

%%%%%%%%%%%%%%%%%%%%%%%%%%%%%%%%%%%%%%%%%%%%%%%%%%%%%%%%%%%%%%%%%%%%%%%%%
% Abstract.        %
%%%%%%%%%%%%%%%%%%%%%%%%%%%%%%%%%%%%%%%%%%%%%%%%%%%%%%%%%%%%%%%%%%%%%%%%%
\begin{abstract}

Nbody simulations are used to examine the consequences of
Neptune's outward migration into the Kuiper Belt, with the 
simulated endstates being compared rigorously and quantitatively
to the observations. 
These simulations confirm the findings of \cite{Cetal03},
who showed that Neptune's migration into a previously stirred--up
Kuiper Belt can account for the Kuiper Belt Objects
(KBOs) known to librate at Neptune's 5:2 resonance. We also
find that capture is possible at many other weak, 
high--order mean motion resonances, such as the
11:6, 13:7, 13:6, 9:4, 7:3, 12:5, 8:3, 3:1, 7:2, and the 4:1.
The more distant of these resonances, such as the 9:4, 7:3, 5:2, 
and the 3:1, can also capture particles in stable, eccentric orbits 
beyond 50 AU, in the region of phase space conventionally known as the
Scattered Disk. Indeed, $90\%$ of the simulated particles 
that persist over the age of the Solar System in the so--called 
Scattered Disk zone never had a close encounter with Neptune, but 
instead were promoted into these eccentric orbits by Neptune's 
resonances during the migration epoch. This indicates that the observed 
Scattered Disk might not be so scattered. This model also produced
only a handful of Centaurs, all of which originated at
Neptune's mean motion resonances in the Kuiper Belt. However
a noteworthy deficiency of the migration model considered here
is that it does not account for the observed abundance of Main Belt KBOs
having inclinations higher than $15^\circ$.

In order to rigorously compare the model endstate with the observed 
Kuiper Belt in a manner that accounts for telescopic selection effects,
Monte Carlo methods are used to assign sizes and magnitudes to the
simulated particles that survive over the age of the Solar System.
If the model considered here 
is indeed representative of the outer Solar System's early
history, then the following conclusions are obtained:
({\it i.})\ the observed 3:2 and 2:1 resonant populations 
are both depleted by a factor of $\sim20$ relative to model expectations; 
this depletion is likely due to unmodeled effects, possibly perturbations
by other large planetesimals, 
({\it ii.})\ the size distribution of those KBOs inhabiting the 3:2 
resonance is significantly shallower than the Main Belt's size distribution, 
({\it iii.})\ the total number of KBOs having radii $R>50$ km and orbiting 
interior to Neptune's 2:1 resonance is $N\sim1.7\times10^5$; these bodies 
have a total mass of $M\sim0.08(\rho/\mbox{1 gm/cm$^3$})(p/0.04)^{-3/2}$ 
M$_\oplus$ assuming they have a material density $\rho$ and an albedo $p$. 
We also report estimates of the abundances and masses of the Belt's various 
subpopulations (e.g., the resonant KBOs, the Main Belt, and the
so--called Scattered Disk), and also provide upper limits on the abundance of
Centaurs and Neptune's Trojans, as well as upper limits on the sizes and abundances
of hypothetical KBOs that might inhabit the $a>50$ AU zone.

\end{abstract}


%%%%%%%%%%%%%%%%%%%%%%%%%%%%%%%%%%%%%%%%%%%%%%%%%%%%%%%%%%%%%%%%%%%%%%%%%
% Introduction.        %
%%%%%%%%%%%%%%%%%%%%%%%%%%%%%%%%%%%%%%%%%%%%%%%%%%%%%%%%%%%%%%%%%%%%%%%%%
\section{Introduction}
\label{introduction}

The Kuiper Belt is the vast swarm of small bodies that inhabit the outer
Solar System beyond the orbit of Neptune. These Kuiper Belt Objects (KBOs)
that inhabit this Belt are relics of the solar system's primordial
planetesimal disk---they are bits of debris that failed to coalesce 
into other large planets. 
The Kuiper Belt is also of great interest since it preserves a record
of the outer Solar System's early dynamical history. This is reflected in
the KBOs' curious distribution of orbits, which
suggest that there was considerable readjustment of the Solar System's 
early architecture. The possibility that the orbits of the giant planets 
may have shifted significantly
(that is, after the solar nebula gas had already dissipated) was
first demonstrated by the accretion simulations of \cite{FI84};
they showed that as the growing giant planets gravitationally scatter
the residual planetesimal debris, they can exchange angular momentum
with the debris disk in a way that causes the planets' orbits to drift.
\cite{M93} later showed that an episode of outwards migration by Neptune
by at least $\Delta a\simeq5$ AU
could also account for Pluto's peculiar orbit, which resides at Neptune's
3:2 resonance with an eccentricity of $e\simeq0.25$. In this scenario,
Pluto's large eccentricity is a consequence of it having been captured by
Neptune's advancing 3:2 resonance, which pumped Pluto's $e$
up as it shepherded the small planet outwards.
Further support for this 
planet--migration scenario is provided by the subsequent discovery of 
numerous other
KBOs also inhabiting Neptune's 3:2 resonance with eccentricities 
similar to model predictions \citep{M95}, as well as by more modern Nbody 
simulations of the orbital evolution of giant planets while they are
still embedded in a massive planetesimal disk (\citealt{HM99, GML04}).
 
The purpose of the present work is to use higher--resolution
simulations to update this conventional model of Neptune's
migration into the Kuiper Belt.
This model's strengths, as well as its weaknesses, will be assessed 
quantitatively by rigorously 
comparing the simulations' endstates to current observations of the Belt. 
In the following, we execute two simulations that track the orbital evolution
of the four migrating giant planets plus $10^4$ massless
test particles (the latter representing the KBOs) over the age of the
Solar system. In one simulation the initial state of the Kuiper Belt 
is dynamically cold ({\it i.e.}, the particles have initial eccentricities and
inclinations of $e$ and $\sin i\sim0.001$), while the second simulation
is of a Kuiper Belt that is initially stirred--up a modest amount
({\it i.e.}, $e$ and $\sin i\sim0.1$). We then use a Monte Carlo method
to assign sizes (and hence magnitudes) to the
simulated KBOs; this allows us to account for the telescopic biases
that tends to select those KBOs that inhabit orbits that are more
favorable for discovery over those KBOs in less favorable orbits. 
Then, by comparing the resulting model Kuiper Belts with the current
observational data, we rigorously test the planet--migration scenario as 
well as obtain a more realistic assessment of the abundance of KBOs.
This analysis will also provide
the relative abundance of the Belt's various subpopulations---the 
resonant KBOs, the Main Belt Objects, the Scattered Disk, plus the 
Centaurs and Neptune's Trojans. 

The paper is organized as follows. Section \ref{model} describes the 
so--called `standard model' that is
considered here, as well as the numerical methods to be employed. 
Results from two simulations of the Kuiper Belt are reported in Sections 
\ref{cold} and \ref{hot}, while Section \ref{i} 
examines the Kuiper Belt inclination problem. 
Section \ref{census} details the Monte Carlo model that is then used in 
Sections \ref{edge}--\ref{extendedSD} to assess the relative abundance of 
the Belt's various subpopulations, with a final tally of the Belt's total 
population given in Section \ref{calibration}. Section \ref{unmodeled} 
comments on some important unmodeled effects, and
Section \ref{discussion} summarizes the results.

%%%%%%%%%%%%%%%%%%%%%%%%%%%%%%%%%%%%%%%%%%%%%%%%%%%%%%%%%%%%%%%%%%%%%%%%%
% Simulating migration.       %
%%%%%%%%%%%%%%%%%%%%%%%%%%%%%%%%%%%%%%%%%%%%%%%%%%%%%%%%%%%%%%%%%%%%%%%%%
\section{Simulating planet migration}
\label{model}

The MERCURY6 Nbody integrator \citep{C99} is used to
track the orbital evolution of the four giant planets plus
numerous massless particles. In our simulations,
planet migration is implemented by applying an external torque to
each planet's orbit so that its semimajor axis $a_j$ varies as 
\begin{equation}
  \label{a_j(t)}
  a_j(t)=a_{f,j}-\Delta_je^{-t/\tau},
\end{equation}
where $a_{f,j}$ is planet $j$'s final semimajor axis, $\Delta_j$ is the
planet's net radial displacement, and $\tau$ is the e--fold timescale
for planet migration; this form of planet migration was first used in
\cite{M93}. To implement this in MERCURY6, the integrator is modified so that 
each planet's velocity $v_j$ is incremented by the small velocity kick
\begin{equation}
  \Delta v_j=\frac{1}{2}\frac{\Delta_j}{a_j}
    \frac{\Delta t}{\tau}e^{-t/\tau}v_j
\end{equation}
with each timestep $\Delta t$. This additional velocity kick is directed
along the planet's velocity vector, and results in a
torque $T_j=m_ja_j\Delta v_j/\Delta t$ being applied to each planet.
Since $T_j=dL_j/dt$ where $L_j=$ the planet's angular momentum, these
velocity kicks cause the planet's orbit to vary at the rate
$\dot{a}_j=2a_jT_j/L_j=(\Delta_j/\tau)e^{-t/\tau}$, which then
recovers Eqn.\ (\ref{a_j(t)}) when integrated.

The simulations reported below adopt the current planets' masses and orbits
as initial conditions, except that their initial semimajor axes are 
displaced by an amount $-\Delta_j$ so that the migration torque 
ultimately delivers these planets into orbits 
similar to their present ones.
The free parameters that describe this migration
are the planets' radial displacements $\Delta_j$
and the migration timescale $\tau$. 
At present, there is only one strong constraint on the
$\Delta_j$'s, namely, that Neptune's orbit must expand by 
$\Delta_N\simeq8.7$ AU if resonance trapping is to account for the KBOs 
having eccentricities of $0<e\lesssim0.32$ at Neptune's 3:2
resonance (see Appendix \ref{appendix_BI}). 
Another constraint, on the magnitude
of Jupiter's inward migration, can be obtained from the orbital distribution
of asteroids in the outer asteroid belt.  \cite{LM97} show
that the severe depletion of the outer asteroid 
belt can be explained if Jupiter migrated inward by at least 0.2 AU, and 
Franklin et al (2004) show that the orbits of the Hilda 
asteroids at Jupiter's 3:2 resonance are consistent 
with Jupiter having migrated inwards by about 0.45 AU.
The remaining $\Delta_j$'s for Saturn and
Uranus are less well--constrained, but stability considerations require them
to be neither too large nor too small. With this in mind, our simulations adopt the
following values for the $\Delta_j$'s:
$\Delta_J=0.900$ AU for Jupiter, $\Delta_S=2.09$ AU for Saturn,
$\Delta_U=4.52$ AU for Uranus, and $\Delta_N=8.70$ AU for Neptune. 
All of the simulations reported here also employ a planet--migration timescale 
of $\tau=10^7$ years. This value is supported by the self--consistent 
Nbody simulations by \citet{HM99} of the giant planets' migration while
they are embedded in a planetesimal disk.  Those simulations show that
a planetesimal disk having a mass $M_D\sim50$ M$_\oplus$ 
spread over $10<a<50$ AU will cause 
Neptune's orbit to expand $\Delta a\sim7$ AU over a characteristic 
timescale of $\tau\sim10^7$ years (see also \citealt{GML04}).

We note that the orbital evolution adopted here is constructed so that
the migrating planets' eccentricities are always comparable to their present values,
and that the migration proceeds along nearly circular orbits. 
But this particular choice for the planets' eccentricities
is merely a simplifying assumption since we do not know the
$e$-evolution of the giant planets during the migration epoch. For instance,
it is  possible that dynamical friction with the particle disk
would have conspired to keep the planets' eccentricities low, 
but there may also have been other transient protoplanets roaming about the outer 
Solar System, and their perturbations would tend to pump up the planets' 
eccentricities. Given the uncertainty in the relative rates of these
effects, we adopt the simplest possible model, one that assumes
that the giant planets' eccentricities were always comparable to their
present one. However alternate migration schemes are possible; for instance, 
\cite{TGML05} consider a scenario where the giant planets
pump up their eccentricities as they pass through mutual resonances. But it is
uncertain as to whether this possible history would have altered the bulk properties
of the Kuiper Belt, and it is not considered here.

In order to enforce migration in nearly circular orbits,
our simulations have Jupiter migrating outwards
$\Delta_J=0.9$ AU, whereas other self--consistent simulations show that 
Jupiter usually migrates inwards a small amount \citep{HM99}. 
Note that this choice avoids having Jupiter approach the 5:2 resonance with 
Saturn, which tends to excite the planets' eccentricities above current levels.
But that eccentricity excitation might then have been damped back to current levels
by dynamical friction with the particle disk, but that is a phenomena that goes 
unmodeled in our massless particle disk. We simply avoid this event
altogether by instead having Jupiter migrate outwards a modest amount. 
But this not a concern here
since our interest is in the Kuiper Belt, whose endstate is not likely to preserve any 
memory of whether Jupiter migrated slightly inwards or outwards. The remaining 
$\Delta_j$'s are similarly chosen to avoid all major resonances, and the 
simulated planets' semimajor axes $a_j(t)$  are also shown in Fig.\ \ref{migrate}. 

\placefigure{migrate}

In the simulations described below,
the model Kuiper Belt is initially composed of $10^4$ massless particles
having semimajor axes randomly distributed over $20<a<80$ AU
with a surface number density that varies as $a^{-2}$. The inner
edge of this particle disk is 1.4 AU inwards of Neptune's initial semimajor
axis, and the disk extends well beyond the outer reaches of the
observed Main Belt. 
All simulations described here use a timestep of $\Delta t=0.5$ years,
which is sufficiently short to accurately evolve particles in eccentric
orbits down to perihelia as low as 
$q_{min}\simeq(10\Delta t/\mbox{1 yr})^{2/3}\mbox{ AU}\simeq3$ AU
without suffering the perihelion instability\footnote{This difficulty is
overcome by the algorithm of \cite{LD00}.} described in \cite{RM99}.
Of course, particles can still achieve orbits having perihelia lower than
$q_{min}$, but their orbits will not be calculated correctly in our
simulations.  However this is of little consequence since
these planet--crossing bodies have very short dynamical lifetimes 
and are quickly removed from the system anyway. 

%%%%%%%%%%%%%%%%%%%%%%%%%%%%%%%%%%%%%%%%%%%%%%%%%%%%%%%%%%%%%%%%%%%%%%%%%
% Cold Kuiper Belt.       %
%%%%%%%%%%%%%%%%%%%%%%%%%%%%%%%%%%%%%%%%%%%%%%%%%%%%%%%%%%%%%%%%%%%%%%%%%
\section{Migration into a dynamically cold Kuiper Belt}
\label{cold}

Accretion models have shown that the observed KBO population must have 
formed in an environment that was initially dynamically cold, that is,
the known KBOs must have formed from seeds that were in nearly circular 
and coplanar orbits with initial $e$'s and $\sin i$'s that were 
$\lesssim10^{-3}$ \citep{S96, KL99}. In anticipation of this, many
investigations of the dynamical history of the Kuiper Belt have
adopted initial KBO orbits that are dynamically cold 
(e.g., \citealt{M93,M95,DLB95,YT99,CJ02}).

\placefigure{cold belt}

Figure \ref{cold belt} shows the results of a simulation
of Neptune's smooth migration into a dynamically cold swarm of
massless Kuiper Belt objects having initial $e$'s that are
Rayleigh distributed about a mean value $\langle e\rangle=0.001$, and initial 
inclinations similarly distributed with a mean
$\langle\sin i \rangle = \langle e\rangle/2$. This system is evolved for
$t=5\times10^8$ years. As is well known from previous studies, 
Neptune's smooth migration is very efficient at inserting particles 
into the planet's mean--motion resonances, 
principally the 2:1, 5:3, and the 3:2.  
However, it is also well recognized that the endstate of this idealized 
model differs from the observed KBO orbits 
(the red dots in Fig.\ \ref{cold belt}) in several ways. For example,
one prominent discrepancy is that the 2:1 resonance is densely populated with
simulated particles while only sparsely populated by observed KBOs. 

However the discrepancy that is most important to this discussion
lies in the $44<a<47$ AU zone between the 7:4 and the 2:1
resonances, which is the outer half of the Main Belt that is conventionally
defined as the region between the 3:2 and 2:1 resonances.
Although these simulated particles managed to slip through the advancing 2:1
resonance, they still reside in orbits that are only modestly disturbed
with $e\sim0.05$ and $i\sim0.5^\circ$, whereas the observed KBOs inhabit
orbits that are considerably more excited. Thus
Neptune's smooth migration into a dynamically cold Kuiper Belt is unable to 
account for the Belt's stirred--up state.

Evidently, some other event has also disturbed the Kuiper Belt, and
this stirring event may have taken place prior to or after the onset of 
Neptune's migration. However Section \ref{hot} provides reason to believe
that this stirring event occurred before the onset of Neptune's migration
into the Kuiper Belt.

%%%%%%%%%%%%%%%%%%%%%%%%%%%%%%%%%%%%%%%%%%%%%%%%%%%%%%%%%%%%%%%%%%%%%%%%%
% Hot Kuiper Belt.       %
%%%%%%%%%%%%%%%%%%%%%%%%%%%%%%%%%%%%%%%%%%%%%%%%%%%%%%%%%%%%%%%%%%%%%%%%%
\section{Migration into a stirred--up Belt}
\label{hot}

To examine the effects of Neptune's migration and its resonance sweeping
of a previously stirred--up Kuiper Belt population, we repeat
the numerical integrations with $10^4$ simulated KBOs, 
but with initial $e$'s Rayleigh distributed about a 
mean value of $\langle e\rangle=0.1$ and
initial $i$'s distributed similarly about a mean value of
$\langle\sin i \rangle = \langle e\rangle/2$.
However this time the simulation is evolved for the age of the Solar 
System, 4.5 Gyrs, with Fig.\ \ref{hot belt} showing the resulting
Kuiper Belt endstate.

\placefigure{hot belt}

First, we note that in this case we find an outer Solar System that is far 
more depleted in transient particles like Centaurs (which are scattered 
particles having semimajor axes interior to Neptune) and Scattered Disk
Objects [which are particles that were lofted into eccentric Neptune--crossing
orbits due to a close--encounter with Neptune \citep{DL97}]; those bodies 
usually reside in orbits having perihelia $q$ between the $q=30$ and $q=40$ AU
curves seen in Figs.\ \ref{cold belt}---\ref{hot belt}.
This difference is primarily due to the simulation's
longer integration time.

Another prominent difference with the `cold belt' simulation is that Neptune's 
weaker higher--order resonances, such as the 3:1 and 5:2, are considerably 
more efficient at capturing particles when Neptune migrates into a hot disk,
a phenomenon that was first noted by \cite{Cetal03}. 
This result was rather surprising, because low-order resonance
capture theory theory predicts a generally lower capture probability for particles
having higher eccentricities \citep{BG84,M93a}. 
However, a careful examination of the theory of adiabatic resonance capture
(e.g., \citealt{DMM88}) shows that there are two reasons for this result.
(1) The higher order resonances have capture probabilities that drop 
off more slowly with eccentricity than first order resonances:
although the 1st order resonance capture probability varies as $\sim e^{-3/2}$,
the 2nd order resonance capture probability falls off more slowly
as $\sim e^{-1}$ while the 3rd order resonance capture probability
varies as $\sim e^{-1/2}$.
(2) The threshold migration speeds for adiabatic resonance capture
are also lower for the higher 
order resonances, and they also depend more strongly upon the 
initial eccentricities. For capture at a $j+k:j$ resonance, the requirement 
for adiabatic resonance capture is that Neptune's migrate rate,
which is $\dot{a}_N\sim\Delta_N/\tau\sim10^{-6}$ AU/yr in these simulations, 
be sufficiently slow, namely, that
\begin{equation}
  |\dot{a}_N|\ll 8jC_{jk}e^k\frac{m_N}{M_\odot}\frac{a_N}{P_N}.
\end{equation}
where $m_N$ and $P_N$ are Neptune's mass and orbital period, and
$C_{jk}$ is a function of Laplace coefficients. 
For example, the 5:2 resonance has $C_{23}\simeq3.3$, so the 
migration speed threshold that permits adiabatic resonance sweeping is
$|\dot{a}_N|\ll5\times10^{-7}$ AU/yr among particles having $e\sim0.1$, 
while the threshold is reduced to
$|\dot{a}_N|\ll5\times10^{-13}$ AU/yr among particles having $e\sim0.001$.
It is clear then that a dynamically cold particle swarm 
has no chance of adiabatic capture at Neptune's high--order mean--motion
resonances, while particles that are stirred up to
$e\sim0.1$ are at least near the threshold for adiabatic resonance capture.  
And as \cite{Cetal03} point out, the fact that seven eccentric KBOs are known 
to librate at Neptune's 5:2 resonance also lends support to the 
pre-stirred Kuiper Belt scenario.

Another advantage of this stirred--up Kuiper Belt scenario is that it recovers
eccentricities that are observed to be as large as  $e\sim0.2$ in
the Main Belt that lies between the
3:2 and 2:1 resonances at $40\lesssim a\lesssim48$ AU
(the red dots in Fig.\ \ref{hot belt}). This is a feature
that the cold Belt scenario (Fig.\ \ref{cold belt}) does not account for.
Of course, Fig.\ \ref{hot belt} also shows that the simulated Main Belt is
densely populated by low--eccentricity particles having
$e\sim0.05$ at $a\sim47$ AU, whereas the observed Kuiper Belt
is only sparely populated here. But Section \ref{edge} will show that
there are a variety of possible explanations for this discrepancy---such a change
in the KBO size distribution, or perhaps an
outer edge in the primordial Kuiper Belt.

Figure \ref{hot belt} also shows that trapping at the distant high--order
resonances like the 5:2 and 3:1 is quite effective at promoting
bodies into eccentric orbits having $a \gtrsim 50$ AU and 
perihelia $30\lesssim q\lesssim 40$ AU.
This domain is usually regarded as the Scattered Disk.
This result then suggests the possibility 
that some of the observed KBOs in the
$30\lesssim q\lesssim 40$ AU zone may actually be resonantly trapped bodies 
that are masquerading as members of the Scattered 
Disk. Of course, particles scattered by Neptune also tend to spend a large 
fraction of their time near resonances due to the resonance sticking 
phenomenon (e.g., \citealt{DL97, MT99}).
Therefore, the discrimination between scattered and resonantly 
trapped particles must be done carefully.
Towards this end, we examine the orbital histories of all surviving
particles in the shaded zone in Fig.\ \ref{sd} that have
$e>0.25$ and $a>50$ AU. The usual definition of being `in' a 
$j+k:j$ mean motion resonance is that the particle's resonance angle 
$\phi_{jk}$, Eqn.\ (\ref{phi_jk}), librates about some fixed value with 
some modest amplitude $|\Delta\phi_{jk}|$ that is usually $\lesssim90^\circ$,
while the resonance angle for a scattered body that is temporarily `stuck' in
a resonance will have a $\phi_{jk}$ that circulates over $\pm180^\circ$.
However, we find that $|\Delta\phi_{jk}|$ is not the best discriminant for 
identifying trapped and scattered particles because a small but significant
fraction of particles do get trapped at a resonance with a $\phi_{jk}$
that is either circulating or else librating with a very large amplitude.
For some trapped particles, this distinction is unclear due to this 
simulation's infrequent time--sampling that occurs every $\Delta t=10^8$ years.

\placefigure{sd}

Rather, a more reliable discriminant between trapped and 
scattered particles is based on Brouwer's integral $B$, Eqn.\ (\ref{B}).
This integral is conserved by resonantly trapped particles but is not 
conserved by scattered particles that are temporarily exhibiting the
`resonance sticking' phenomenon.  Of the 134 particles
that inhabit the gray zone in Fig.\ \ref{sd}, only 12, or about $10\%$
of these particles, are truly scattered particles whose orbits $(a,e)$
evolve stochastically about the $30\lesssim q\lesssim 40$ AU zone; 
these scattered particles are indicated by the crosses in Fig.\ \ref{sd}. 
The remaining particles are resonantly trapped particles, 
most\footnote{However the $e$'s and
$i$'s of some resonantly trapped particles will still oscillate with constant $a$
in a manner that preserves their Jacobi integral; this evolution usually occurs
after migration has ceased, and these particular 
motions do {\em not} preserve $B$.}
of which preserved their $B$ integral to within $\sim3\%$.

The orbits of all particles having perihelia $q<40$ AU have also been inspected,
and those resonances inhabited by trapped particles having libration amplitudes
$|\phi_{jk}|\le90^\circ$ are indicated by the vertical dashes in the Figure. 
We find that particles get trapped at a number of exotic resonances like the 11:6,
13:7, 13:6, 9:4, 12:5, 8:3, and the 11:4. 

%%%%%%%%%%%%%%%%%%%%%%%%%%%%%%%%%%%%%%%%%%%%%%%%%%%%%%%%%%%%%%%%%%%%%%%%%
% Kuiper Belt Inclinations.      %
%%%%%%%%%%%%%%%%%%%%%%%%%%%%%%%%%%%%%%%%%%%%%%%%%%%%%%%%%%%%%%%%%%%%%%%%%
\section{Kuiper Belt inclinations}
\label{i}

Inspection of the inclinations shown in Fig.\ \ref{hot belt} suggests that
the smooth migration model does not produce sufficient
numbers of bodies in high--inclination orbits.
This has been recognized in previous studies (\citealt{M95, G97}),
but has not been quantified. However one should not directly compare the
simulation's $i$'s to the observed KBO inclinations, since the latter
is heavily biased by telescopic selection effects. Note that
most telescopic surveys of the Kuiper Belt observe near the ecliptic, which 
favors the discovery of lower--$i$ KBOs that spend a larger
fraction of their time at lower latitudes \citep{JL95}. 
To mitigate this selection effect,
one should instead consider the ecliptic inclination distribution, 
which is the inclination distribution of objects having
latitudes $\beta$ very near the ecliptic (e.g., \citealt{B01}). The 
ecliptic inclination distribution for multi--opposition KBOs having 
perihelia $q\le42$ AU and latitudes $|\beta|\le1^\circ$ (red curve)
is shown in Fig.\ \ref{idist}, 
as well as the simulated ecliptic inclination distribution for particles from
Fig.\ \ref{hot belt} that are selected similarly (black curve). 

\placefigure{idist}

The agreement seen in 
Fig.\ \ref{idist} among bodies having inclinations of 
$0\lesssim i\lesssim15^\circ$
shows that the smooth migration model can readily recover the Kuiper Belt's 
lower inclination members. Of course, this
agreement is partly due to the particles' initial inclinations being 
distributed around $\langle i \rangle\simeq3^\circ$. 
But Fig.\ \ref{idist} also shows this model to be
quite deficient in producing sufficient numbers of the
high--$i$ bodies having $i\gtrsim15^\circ$. 
Similar results are also obtained among bodies orbiting at Neptune's 3:2 resonance.
This is a serious discrepancy, since
\cite{B01} has shown that there are two inclination--populations in the Kuiper 
Belt: a minor population of low--$i$ having characteristic inclinations of 
$i\sim3^\circ$, and a high--$i$ population having $i\sim15^\circ$ containing
about three--quarters of all KBOs. Note that these high--$i$ bodies are very 
underrepresented in Fig.\ \ref{idist} due to telescopic selection effects.

Of course, Neptune--scattered  particles 
routinely achieve high inclinations of $i\gtrsim15^\circ$; for instance, 
many of the high--$i$ particles seen in
Fig.\ \ref{cold belt} were scattered by Neptune. Could 
the Scattered Disk be a source of the high--$i$ KBOs
that are found elsewhere in the Belt? 
Recent Nbody simulations by \cite{G03} show that a small fraction
of these Neptune--scattered particles can evolve
from very eccentric, Neptune--crossing orbits into less eccentric orbits in the Main
Belt. In Gomes' simulations, this occurs
principally at secular and mean motion
resonances that drive large oscillations in a scattered particle's
eccentricity. When a scattered particle visits a resonance, 
it can have its $e$ temporarily lowered and its $q$ raised.  If this occurs
during the planet--migration epoch, this process becomes irreversible and can
strand Scattered particles in the Main Belt 
with their high inclinations. Such bodies are identified by Gomes
as `evaders' since they are Neptune--crossing bodies that ultimately
manage to evade Neptune when deposited in the Main Belt.
Note, however, that
the efficiency of this process is quite low, affecting only
$\epsilon\sim0.002$ of all Scattered particles in the simulation
that is evolved over the age of the Solar System by \cite{G03}. 
However all of the high--$i$ particles seen in our simulation 
(Fig. \ref{hot belt}) achieved their inclinations 
while  temporarily or permanently
trapped in Neptune's advancing resonances. 
There were no Neptune--scattered evaders 
having $i>10^\circ$ that survived in our simulations.

Despite the evader mechanism's inefficiency, a model can still be constructed
that yields a KBO inclination distribution that is quite
similar to the observed one. For instance, this can be achieved by making the 
number density of small bodies
initially orbiting interior to $\sim27$ AU about 60 times higher
than the density of bodies initially orbiting beyond 27 AU.  As \cite{G03} show,
Neptune's migration through this densely--populated inner disk creates
the Kuiper Belt's high--$i$ evaders, while the sparse outer disk provides the 
Belt's low--$i$ component. Although this scheme yields an $i$--distribution 
that does indeed agree with the observations, that success is achieved via a special 
configuration of the initial particle disk. 

However, \cite{LM03} avoid this
problem of special initial conditions by assuming
that the initial planetesimal disk simply ended at $\sim35$ AU.  This is the
`push--out' model, which argues that most of the Kuiper Belt is a consequence of
Neptune's advancing 2:1 resonance, which can drag 
bodies outwards to litter the Main Belt with low--$i$ KBOs. The Belt's
high--$i$ component is then presumed to be due to the evader mechanism.
While the push--out model remains quite intriguing, 
it would be interesting to see this scenario subjected to greater scrutiny
to see whether it can indeed reproduce the Kuiper Belt's
curious mix of high and low inclination KBOs in a self--consistent manner.

%%%%%%%%%%%%%%%%%%%%%%%%%%%%%%%%%%%%%%%%%%%%%%%%%%%%%%%%%%%%%%%%%%%%%%%%%
% Census.        %
%%%%%%%%%%%%%%%%%%%%%%%%%%%%%%%%%%%%%%%%%%%%%%%%%%%%%%%%%%%%%%%%%%%%%%%%%
\section{A Kuiper Belt census: comparison with observations}
\label{census}

Figure \ref{abundance} plots the relative abundance, over time, of the 
simulated Belt's various dynamical classes 
among particles having perihelia $q\le 45$ AU. These curves
are normalized such that the final abundance of the Main Belt (MB), where 
$40.1\le a\le 47.2$ AU, is unity. 
Note that this model predicts a 2:1 resonance that is 1.4 times more
abundant than the Main Belt, and 2.5 times more abundant than the 3:2,
while the observations (Fig.\ \ref{hot belt})
show a 2:1 that is only sparsely populated. Of course, when comparing
the simulated population to the observed population, one must first deal
with the observational selection effect that strongly favors the discovery
of larger and/or nearer KBOs. However it is shown below that the
effects of this bias can be accounted for by using a Monte Carlo
method that assigns random sizes to the simulated population. This then allows 
one to make a fair comparison of the relative abundances of the simulated and 
observed populations. 

\placefigure{abundance}

Begin by letting $N(R)=$ the number of bodies in the simulated population
having radii exceeding a radius $R$. 
Also let $\xi$ be a random number that is uniformly distributed
between zero and one, and interpret this number as the
probability of selecting a body with a radius that is smaller than $R$. 
This is also equal to the probability of {\em not} selecting a body
of radius $>R$, so $\xi=1-N(R)/N_{total}$ where $N_{total}$ is the total
number of bodies in this population. Since most small--body populations
have a cumulative size distribution $N(R)$ that varies
as a power law, adopt
\begin{eqnarray}
  \label{N(R)}
  N(R)&=&N_{total}\left(\frac{R}{R_{min}}\right)^{-Q}
\end{eqnarray}
where $R_{min}$
is the radius of the smallest member of the swarm. 
Then $R/R_{min}=(1-\xi)^{-1/Q}$,
but $1-\xi$ can be replaced with $\xi$ since these random numbers have the same 
distribution, so
\begin{eqnarray}
  \label{Rrandom}
  R(\xi)&=&\frac{R_{min}}{\xi^{1/Q}}.
\end{eqnarray}
Equation (\ref{Rrandom}) is then used to generate random sizes for the simulated 
population of Fig.\ \ref{hot belt} that have apparent R--band magnitudes of
\begin{eqnarray}
  \label{mR}
  m_R=m_\odot-2.5\log\left[p\left(\frac{R}{r_1}\right)^2
    \left(\frac{r}{r_1}\right)^{-2}\left(\frac{r-r_1}{r_1}\right)^{-2}\right]
\end{eqnarray}
where $r$ is the particle's heliocentric distance, $r_1=1$ AU, 
$m_\odot=-27.29$ is the Sun's apparent $R$ magnitude,
$R_{min}=20$ km is adopted here, 
and the observation is presumed to occur at solar opposition.
All of our calculations will also adopt the usual albedo of $p=0.04$ so that our
findings can be readily compared to past results obtained by others. 
However if an alternate albedo $p$ is desired, 
simply revise all KBO radii reported here by a factor of $(p/0.04)^{-1/2}$,
and all masses by a factor of $(p/0.04)^{-3/2}$.
Finally, note that a power--law size distribution results in a cumulative
luminosity function that varies as $\Sigma(m)\propto10^{\alpha m}$, where 
$\Sigma(m)$ is the sky--plane number density of KBOs brighter than apparent magnitude
$m$ and the logarithmic slope is $\alpha=Q/5$ \citep{ITZ95}. 

Hubble Space Telescope
observations reveal that the bright end of the Kuiper Belt's
luminosity function has a steep logarithmic slope of $\alpha=0.88\pm0.1$
for bodies having magnitudes $m_R\ll24$,
while the faint end ($m_R\gg24$) of the luminosity function has a shallow 
logarithmic slope of $\alpha=0.32\pm0.15$ \citep{Bernstein04};
the steeper slope of the bright end of the luminosity function
was also confirmed recently by \cite{Eetal05}.
This luminosity function can be interpreted as evidence that the KBO size
distribution is actually two power laws that break even at
a magnitude $m_{break}\simeq24$, which corresponds to a body
of radius $R_{break}\simeq65$ km orbiting at a characteristic distance
of $r\sim40$ AU assuming it has 
an albedo of $p=0.04$. However our application will 
concentrate only on those KBOs that have known orbits, 
and $99\%$ of those bodies have magnitudes
$m_R<m_{break}$. Consequently, this study will be sensitive only to the larger 
end of the KBO size spectrum, and such bodies will be characterized here via a 
single power--law size distribution having $\alpha=0.88$ and $Q=5\alpha=4.4$.

Although $\sim5000$ simulated particles in Fig.\ \ref{hot belt}
manage to survive over the age of the Solar System,
the Monte Carlo model assigns far too few of them with sizes
that would be detected by any telescopic survey of the Kuiper Belt.
To boost the statistics of the detectable portion of the simulated population,
each survivor is replicated $10^4$ times such that
each particle's orbit elements $a,e,i,\Omega,\omega$ are preserved
while its mean anomaly $M$ is randomly distributed
over $0\le M<2\pi$. It should be noted that this step effectively
assumes that the particles' longitudes are uniformly distributed over $2\pi$,
which is not quite correct since Neptune's resonant perturbations tend to
arrange the particles' longitudes in a non--uniform manner \citep{M96, CJ02}.
Nonetheless, this is not a major concern since
the observed KBOs were discovered along 
lines--of--sight that are roughly distributed uniformly in ecliptic longitude,
which effectively washes--out Neptune's azimuthal arrangement of the Belt;
see Appendix \ref{symmetry} for a more detailed examination and justification 
of this assumption.
Lastly, this Monte Carlo model is then tested by verifying that the 
randomly generated population does indeed exhibit the expected luminosity
function that varies as $\Sigma(m)\propto10^{Qm/5}$.

Further comparison of the Monte Carlo model of the Kuiper Belt to any 
observations must be done carefully. Note that the brighter KBOs tend to be
discovered in shallow, wide--angle surveys that observe a large area 
$\Delta\Omega$ on the sky, while the fainter KBOs tend to be discovered in 
deeper surveys that observe smaller areas $\Delta\Omega$. Consequently, the 
observed abundances of the various KBO
subclasses (e.g., the Main Belt, the Scattered Disk, {\it etc.}) are proportional 
to all of these surveys' total area $\Omega(m_R)$, which itself is 
some function of the limiting magnitude $m_R$. However 
Appendix \ref{appendix_ratios} shows that this dependence upon
$\Omega(m_R)$ can be factored out by constructing
ratios of the Belt's various subclasses.  That Appendix 
also shows that the ratios of the observed
abundance of any two dynamical classes of KBOs is approximately equal to 
the ratio of the intrinsic abundances of the much larger
unseen populations\footnote{Of course, this method
of analyzing the Belt's relative abundances will tells us little about
those KBO populations that are either too rare, dim, or otherwise too difficult to 
recover in telescopic surveys. Nonetheless, we still can use our method
to place upper limits on the abundances of any hypothetical KBO populations
that are unseen using the method described in Section \ref{edge}.}. 
Thus by plotting {\em ratios} of the simulated
populations to the observed KBO populations, we can compare the model to 
the observations in a manner that is insensitive to survey details like their 
individual sky--coverage $\Delta\Omega$.

Figure \ref{ratios} shows the apparent abundances of the 2:1 and the 3:2 
populations relative to the Main Belt (MB) as a function of
their $R$--band magnitudes $m_R$. The upper light gray curve is the
simulated ratio, which predicts an apparent 2:1 abundance that is about $80\%$
of the MB, while the dark gray curve is the observed
ratio. Taking the ratio of these two curves reveals that their discrepancy
at magnitudes $m_R>21$ (which refers to about 90\% of the 
observed sub--populations) is a factor of $f_{2:1}\simeq20$---the 
observed 2:1 resonance is markedly underabundant relative to the observed Main 
Belt population. There are $N_{MB}(R>50\mbox{ km})=1.0\times10^5$ 
Monte Carlo particles in the simulated Main Belt having radii $R>50$ km, and 
$N_{2:1}(R>50\mbox{ km})=8.2\times10^4$ 
particles in the 2:1. If we let $r_{2:1/MB}$ represent the inferred ratio of 2:1 to 
Main Belt objects, then $r_{2:1/MB}\simeq N_{2:1}/f_{2:1}N_{MB}\simeq0.041$,
which is comparable (albeit lower by a factor of $\sim2$) to
the ratio that \cite{TJL01} infer from telescopic surveys of the Kuiper Belt.

\placefigure{ratios}

The observed 3:2/MB ratio plotted in Figure \ref{ratios} also shows that this
resonant population is underabundant relative to model predictions by a factor
of $\sim6$ among bright objects with $21\lesssim m_R\lesssim23$, 
and by a factor of $\sim60$
at fainter magnitudes. Close inspection of the observations suggests that
there indeed is a deficiency of fainter KBOs in the 3:2, 
and that this curve is not due to some overabundance of Main Belt KBOs
having magnitudes of $m_R\simeq23$. 

It should be noted that the results given in Fig.\ \ref{ratios} are not particularly
sensitive to the detailed location of the Main Belt's outer edge. 
For instance, if we assumed the Belt's primordial edge where instead
at $a=45$ AU (e.g., \citealt{TB01}; see also Section \ref{edge}), 
this reduces the 2:1 and MB populations
both by about 40\% while leaving the 3:2 population 
unchanged. Consequently, the 2:1/MB ratios of Fig.\ \ref{ratios}
are largely unchanged for both the simulated and observed populations,
while the 3:2/MB ratios increase by a factor of $(1-0.4)^{-1}\simeq1.7$.
However the {\em discrepancy} between the simulated and observed populations
is still the same factors of $\sim6$--60.

Although there are several possible interpretations of the discrepancies
seen Fig.\ \ref{ratios}, the most plausible explanation is
that other unmodeled processes are responsible for
({\it i.}) reducing the trapping efficiencies of the 2:1 and 3:2 resonances
by factors of $\sim6$--60, or ({\it ii.}) causing trapped particles
to diffuse out of the resonances and into nearby regions of phase space
that are quite unstable (cf.\ Fig.\ 1 of \citealt{DLB95}),
resulting in their ejection from the Kuiper Belt. Such unmodeled processes
include the collisions and gravitational scatterings that occurred with ever 
greater vigor during earlier times when the Belt was more crowded. 
The scattering of these planetesimals by Neptune was of course
responsible for driving that planet's migration,
so the occasional scattering of a large and/or close planetesimal will cause
that planet's orbit and hence its resonances to shudder some. Likewise, 
scattering events among the KBOs themselves would
also cause their semimajor axes to diffuse some, as would
collisions. This means that scatterings and collisions will have
driven a random walk in the resonant particles' semimajor axes, as well as a 
random walk in the location of the resonances themselves. 
It is possible then that these unmodeled effects can drive particles out of 
resonances and reduce the resonant population by the 
large factors indicated by Fig.\ \ref{ratios}, a scenario that is
also explored in simulations by \cite{ZSZZV02} and \cite{TM04}.

The magnitude dependence of the observed 3:2/MB ratio shown in 
Fig.\ \ref{ratios} is also quite curious. The fact that the observed ratio 
varies with apparent magnitude $m_R$, while the simulated ratio remains 
constant at magnitudes fainter than $m_R=21$, 
suggests that the $N(R)\propto R^{-Q}$ power--law that was
universally applied throughout the entire Belt is overly simplistic.
One way for the model to achieve better agreement with the observations
is to assume that larger, brighter bodies are more abundant in the 3:2,
and that smaller, fainter bodies are less abundant there than they are in the MB, 
which requires a shallower size distribution.
The dashed curve in Fig.\ \ref{ratios} illustrates this possibility,
which shows the simulated 3:2/MB ratio assuming that the 3:2 bodies
have a shallow $Q=2.7$ size distribution with $R_{min}=4.3$ km 
(note that reducing $R_{min}$ has the effect of reducing the total number of 
visible objects) while the MB bodies have the usual distribution with $Q=4.4$ 
and $R_{min}=20$ km. The shallow size distribution that is
inferred here for the 3:2
population is also consistent with the logarithmic slope of $\alpha\simeq0.56$
that \cite{Eetal05} recently reported for the luminosity function of
their `resonant' population that is dominated by 3:2 KBOs;
the size distribution inferred from that work is $Q=5\alpha=2.8$. 
KBO sizes can also vary with inclination
[\cite{LS01}, but see also footnote 1 of \cite{G03}]. In particular, \cite{Bernstein04} 
report that the bright end of the luminosity function for
high--inclination ($i>5^\circ$) KBOs have
a logarithmic slope of $\alpha=0.66$ and a size distribution
$Q=5\alpha=3.3$ that is much shallower than the low--$i$ KBOs having
$\alpha=1.36$ and a $Q=5\alpha=6.8$.

In our Monte Carlo model there are only
$N_{3:2}(R>50\mbox{ km})=2100$ bodies larger than $R=50$ km, so their
numerical abundance relative to the Main Belt is
$r_{3:2/MB}=N_{3:2}/N_{MB}=0.021$, which again is comparable
(but again lower by a factor of $\sim2$)
to the ratio reported in \cite{TJL01}.
The largest Monte Carlo body at the 3:2 has a radius
$R\simeq1000$ km, which is comparable to the size of the 
largest multi--opposition
KBO there\footnote{excepting Pluto of course which has an
anomalously high albedo of $p\sim0.5$.} with $R\simeq1100$ km, 
assuming $p=0.04$.

This range of power--law indices that is inferred for the Kuiper Belt, 
$2.7\lesssim Q\lesssim4.4$, is comparable to the values of $Q$ that are
observed at various sites throughout the asteroid belt. The Near Earth 
Objects have a fairly shallow size distribution with $Q=1.95$ \citep{SB04}, 
while  the asteroid families exhibit steeper size distributions. For instance, 
Fig.\ 1.\ of \cite{TCMZPd99} shows size distributions for several prominent asteroid
families having values of $2\lesssim Q\lesssim6$. 
Note also that nonfamily asteroids have $Q\simeq3.0$
\citep{Ietal01}, which is slightly steeper than the canonical $Q=2.5$ value
that results from a collisional cascade \citep{D69}. Since the various
asteroid subclasses exhibit such a wide variation in their size distributions
over a relatively narrow range of semimajor axes of
$\Delta a\sim4$ AU, perhaps it should be of no surprise that the spatially
much wider
Kuiper Belt might also exhibit some variety in $Q$.

%%%%%%%%%%%%%%%%%%%%%%%%%%%%%%%%%%%%%%%%%%%%%%%%%%%%%%%%%%%%%%%%%%%%%%%%%
% Edge.         %
%%%%%%%%%%%%%%%%%%%%%%%%%%%%%%%%%%%%%%%%%%%%%%%%%%%%%%%%%%%%%%%%%%%%%%%%%
\section{The outer edge of the Solar System}
\label{edge}

\begin{figure}
\epsscale{1.0}
\plotone{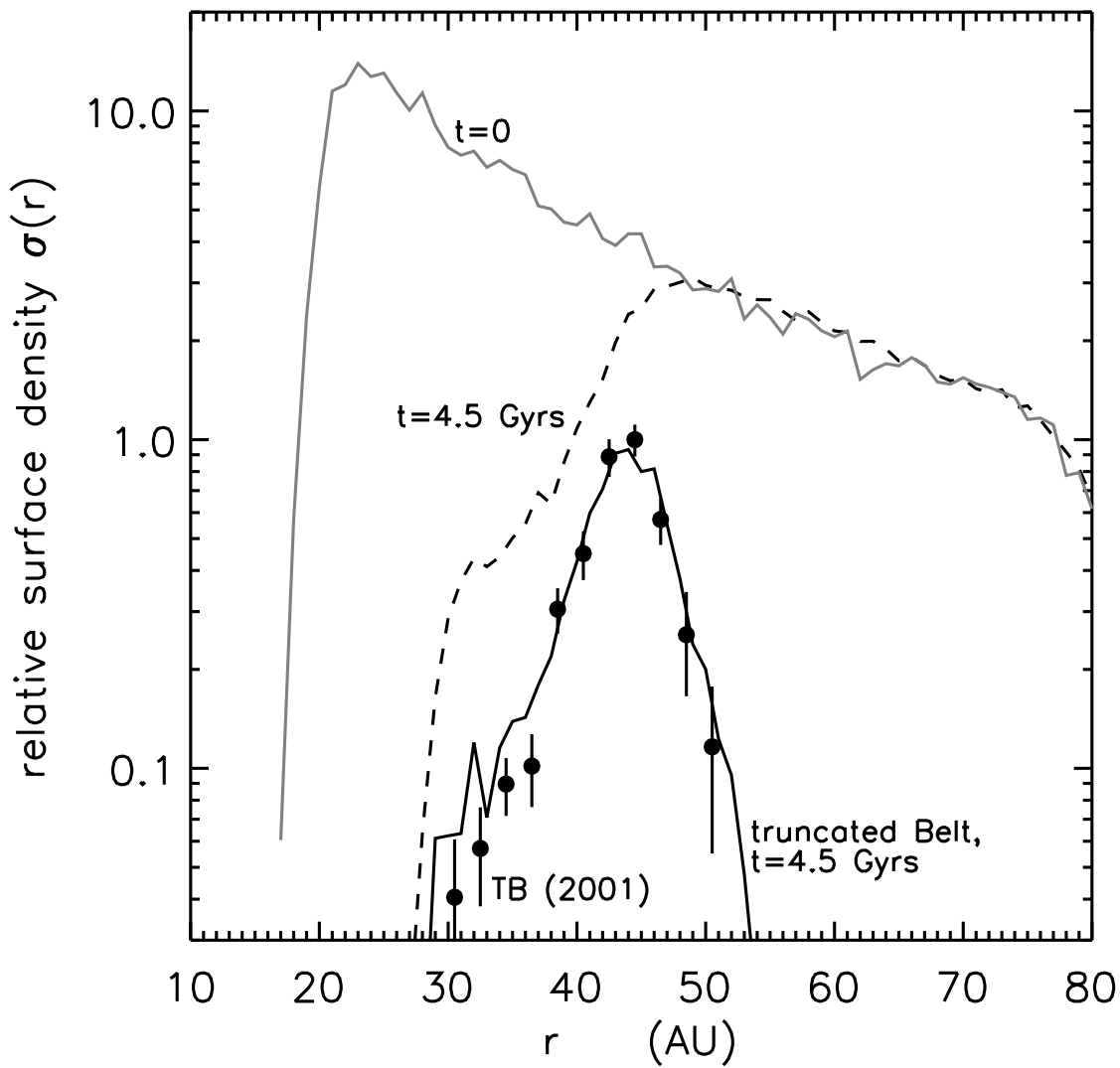}
\figcaption{
  \label{erosion}
  Dots show the KBO surface number density $\sigma(r)$, normalized
  to peak at unity, that is inferred from the KBOs radial distribution function 
  $f(r)\propto r\sigma(r)$ reported by \cite{TB01}.
  The $t=0$ and $4.5$ Gyr curves are the simulated Belt's initial and final
  surface density. The `truncated Belt' curve is the simulation's
  final surface density assuming that the Belt is truncated at $a=45$ AU,
  and that the 3:2 population also does not contribute to $\sigma$.
  Note that the surface density of the inner half of this Belt increases
  as a very steep function of distance as 
  $\sigma(r)\propto r^{9.5}$ for $r\lesssim45$ AU.
}
\end{figure}

Inspection of Figure \ref{hot belt} shows a prominent absence
of observed KBOs having modest eccentricities of $e\sim0.1$
near and beyond Neptune's 2:1 resonance. The prevailing interpretation of
this observed feature is that there is a boundary near $45\lesssim a\lesssim50$ AU
that marks the outer edge of the Solar System's primordial
Kuiper Belt \citep{ABM01,TB01}. The dots in Fig.\ \ref{erosion}
shows the Belt's surface density $\sigma(r)$ that is inferred 
from these observations, which peaks at $r\simeq45$ AU. 
We have nonetheless allowed
our simulated Kuiper Belt to extend out to $a=80$ AU in order to use
the dearth of observed distant KBOs to place
quantitative upper limits on the abundance of hypothetical KBOs that might
live beyond 50 AU.

\placefigure{erosion}

\begin{figure}
\epsscale{1.0}
\plotone{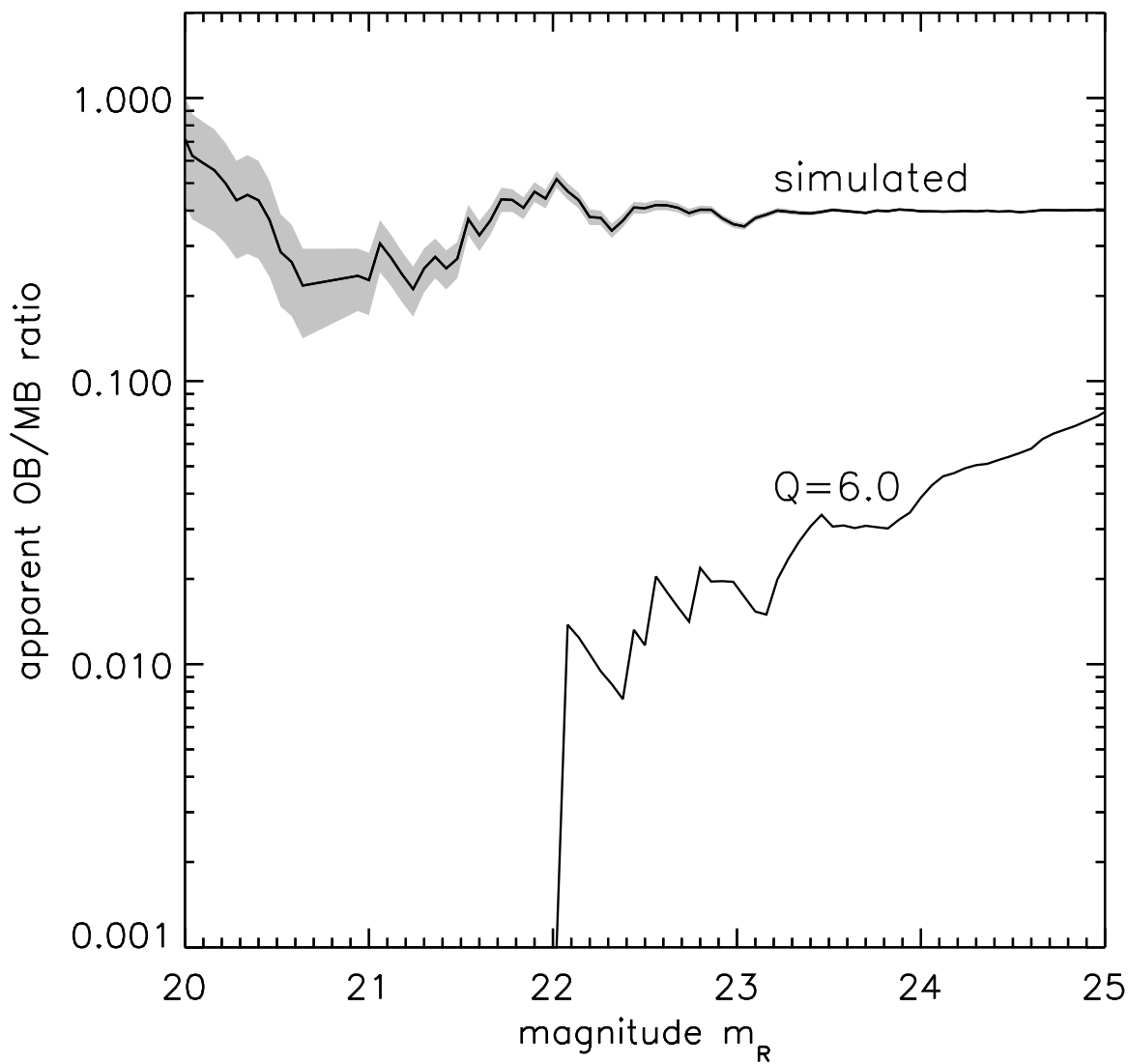}
\figcaption{
  \label{outer_belt}
  The gray curve is the expected ratio of bodies in the Outer Belt
  (OB, beyond $a>50$ AU) relative to the Main Belt
  (MB, $40.1<a<47.2$ AU) as a function of magnitude $m_R$ assuming
  the size distribution has $Q=4.4$ and $R_{min}=20$ km.
  The solid curve is the OB/MB ratio assuming the OB instead has
  a steeper $Q=6.0$ size distribution while the MB has $Q=4.4$,
  with $R_{min}=20$ km for both populations.
}
\end{figure}

The Nbody/Monte Carlo of the previous Section can be used to predict
how many KBOs should have been observed in the $a>50$ AU zone (which we
identify here as the Outer Belt, or OB) assuming
({\it i.})\ that the primordial Kuiper Belt extends smoothly out to 
$a=80$ AU, and ({\it ii.})\ that all KBOs everywhere have the same size 
distribution with the usual parameters $Q=4.4$ and $R_{min}=20$ km.
This simulation's ratio of Outer Belt/Main Belt objects,
$r_{OB/MB}(m_R)$, is plotted versus  magnitude $m_R$ in Fig.\ \ref{outer_belt}.
This is the ratio of $N^{sim}_{OB}(m_R)=$ the number of bodies in the simulated 
Outer Belt (whose members have semimajor axes $50<a<80$ AU and 
eccentricities $e<0.2$), to $N^{sim}_{MB}(m_R)=$ the number of bodies in 
the simulated Main Belt (where $40.1<a<47.2$ AU)
in the magnitude interval $m_R\pm\Delta m$ where $\Delta m=0.5$.
According to the Figure, the expected OB/MB ratio is $r_{OB:MB}=0.4$. At
present there are $N^{obs}_{MB}=264$ KBOs in the MB that have been observed
for 2 or more oppositions, and the dimmest member of this group of KBOs
has an apparent magnitude $m_R^\star=24.5$. The Nbody/Monte Carlo model
thus predicts that there should also
be $N^{obs}_{OB}=r_{OB/MB}N^{obs}_{MB}\simeq100$ objects brighter than 
$m_R^\star$ orbiting in the OB
beyond $a=50$ AU. This prediction is in marked contrast
with the observations which show that there are no known multi--opposition objects 
orbiting in the OB with magnitudes brighter than $m_R^\star$, {\it i.e.},\
$N^{obs}_{OB}<1$. 

\placefigure{outer_belt}

This prediction that the Outer Belt would have an observed abundance that is
$40\%$ of the Main Belt differs considerably from that of \cite{GKNLB98} 
who estimated that the observed
Outer Belt population should only be $\sim6\%$ of the total observed population.
However this much lower estimate was obtained by assuming that the current
Belt's surface number density resembles its primordial $\sigma(r)$ which likely 
varied as $r^{-2}$ or so. This very common assumption causes the inner 
part of the  Belt to be more concentrated than its outer part. 
However a real Kuiper Belt has been dynamically eroded over the eons by the 
giant planets' gravitational perturbations.  This dynamical erosion is illustrated 
in Fig.\ \ref{erosion}, which shows the simulated Belt's
primordial surface density (gray curve) and its final surface density (dashed curve);
similar erosion is also seen in the long--term integrations of \cite{DLB95}.
Figure \ref{erosion} shows that $\sigma(r)$ for an eroded Belt
is a sharply increasing function of $r$ for $r\lesssim50$ AU,
which implies that the inner observable portion of the Belt is very underdense
relative to the more distant $r\sim50$ AU zone. This
dynamical erosion accounts for
the discrepancies between our Outer Belt predictions and that by \cite{GKNLB98}.

Recall that Figs.\ \ref{hot belt} and \ref{ratios} show that the
2:1 and 3:2 populations are very depleted relative to model predictions,
and that the zone beyond Neptune's 2:1 resonance is either empty or inhabited by 
bodies too small and faint to be seen. To account for these depletions,
the solid curve in Fig.\ \ref{erosion} also shows a revised surface density curve 
that is obtained from the simulated Belt that is truncated at
$a=45$ AU (about 3 AU inwards of Neptune's 2:1 resonance),
and with the negligible contribution from the 3:2 populations also
being ignored. This result is a curve that
agrees quite well with the Belt's observed surface density variations.
Despite this good agreement in the radial distributions of the simulated and 
observed Kuiper Belts, Fig.\ \ref{hot belt} shows that this apparent edge
at $a=45$ AU is still rather fuzzy since there are four multi-opposition KBOs
of low eccentricity ($e<0.05$) orbiting in the Main Belt at
 $45<a<48$ AU. Close inspection of Fig.\ \ref{hot belt} shows that a
hard edge at $a=45$ AU also could not account for the KBOs having
$e\sim0.1$ in the $45<a<48$ AU, unless the advancing 2:1 resonance
also dragged some bodies out of the $a<45$ AU zone and deposited them
here, reminiscent of the scenario suggested by \cite{LM03}.

The remainder of this Section places upper limits on the
size and abundance of any unseen KBOs that might lurk beyond $a=50$ AU.
Of course there are multiple interpretations of the 
dearth of observed multiple--opposition bodies orbiting beyond $a=50$ with modest
eccentricities of $e\sim0.1$, {\it i.e.},\ that $N^{obs}_{OB}<1$.
One interpretation of this upper limit
is that assumption ({\it i.})\ is incorrect---that the primordial
Kuiper Belt's density did not extend smoothly beyond Neptune's 2:1, but that it
instead was reduced by a factor $f$ (relative to the smooth model's density)
in the $a>50$ AU zone. In this case, the OB/MB ratio
becomes $r_{OB/MB}=0.4/f=N^{obs}_{OB}/N^{obs}_{MB}<1/N^{obs}_{MB}$,
which implies that the primordial density of the OB was smaller than
the MB by a factor $f\gtrsim100$.

Alternatively, assumption ({\it ii.})\ could be incorrect, namely, that the KBO size
distribution was not uniform everywhere. For instance, the absence
of any multi--opposition bodies in the OB having magnitudes brighter than
$m_R^\star=24.5$ could simply mean that bodies beyond $r\simeq50$ AU
are dimmer than $m_R^\star$ and thus have radii smaller than 
$R\simeq80(p/0.04)^{-1/2}$ km (see Eqn.\ \ref{mR}). Note that
\cite{TJL01} obtained a similar limit, but that they came to regard this scenario
as unlikely.

It is also possible that the Outer Belt's size distribution is steeper,
{\it i.e.},\ has a larger $Q$, than the Main Belt's size distribution.
An increase in $Q$ decreases the abundance of bright bodies, 
as is illustrated by the curve in Fig.\ \ref{outer_belt} which gives the
the $r_{OB/MB}$ ratio for an OB having a $Q=6.0$ size distribution
while bodies in the MB have the usual $Q=4.4$ distribution. 
This particular $Q$ is also the minimum value that
is consistent with the observed upper limit of $r_{OB/MB}<1/264$; Outer Belts with
with a smaller $Q$ would contain at least 1 KBO brighter than $m_R^\star$
in the $a>50$ AU zone for every 264 KBOs detected in the Main Belt, while an OB 
having a larger $Q$ would be undetected. This particular model is near
the threshold of detection, and its largest member 
has a radius of $R=250$ km. \cite{TJL01} also considered this scenario, but
they concluded that the absence of distant KBOs requires a steeper $Q>9$
size distribution. The origin of this discrepancy is unclear.

It is also interesting to note that the low--inclination KBOs have a logarithmic
slope of $\alpha=1.36$ along the bright end of their luminosity function
\citep{Bernstein04}, which implies a steep size distribution of $Q=5\alpha=6.8$.
Such bodies, if they inhabit the Outer Belt beyond $a=50$ AU with the
same abundances as adopted by our model, could conceivably have avoided
detection to date due to their steep size distribution. In other words, a Main Belt
whose low--$i$ population extends beyond $a=50$ AU while its
high--$i$ population terminates at $a=50$ AU could be quite
consistent with their non--detection.

From these considerations it may be concluded that the observed absence of
multi--opposition KBOs in the $a>50$ AU zone having modest eccentricities 
$e\sim0.1$
implies: ({\it i.})\ that this part of the primordial Kuiper Belt was
underdense by a factor of $f\gtrsim100$ relative to the
$a<50$ AU zone, or that ({\it ii.})\ these distant KBOs have radii
$R\lesssim80$ km, or that ({\it iii.})\ their size distribution has
a power--law index $Q>6.0$, or perhaps ({\it iv.}) some combination
of the above effects.

%%%%%%%%%%%%%%%%%%%%%%%%%%%%%%%%%%%%%%%%%%%%%%%%%%%%%%%%%%%%%%%%%%%%%%%%%
% Centaurs.        %
%%%%%%%%%%%%%%%%%%%%%%%%%%%%%%%%%%%%%%%%%%%%%%%%%%%%%%%%%%%%%%%%%%%%%%%%%
\section{The origin of Centaurs}
\label{centaurs}

\begin{figure}
\epsscale{1.0}
\plotone{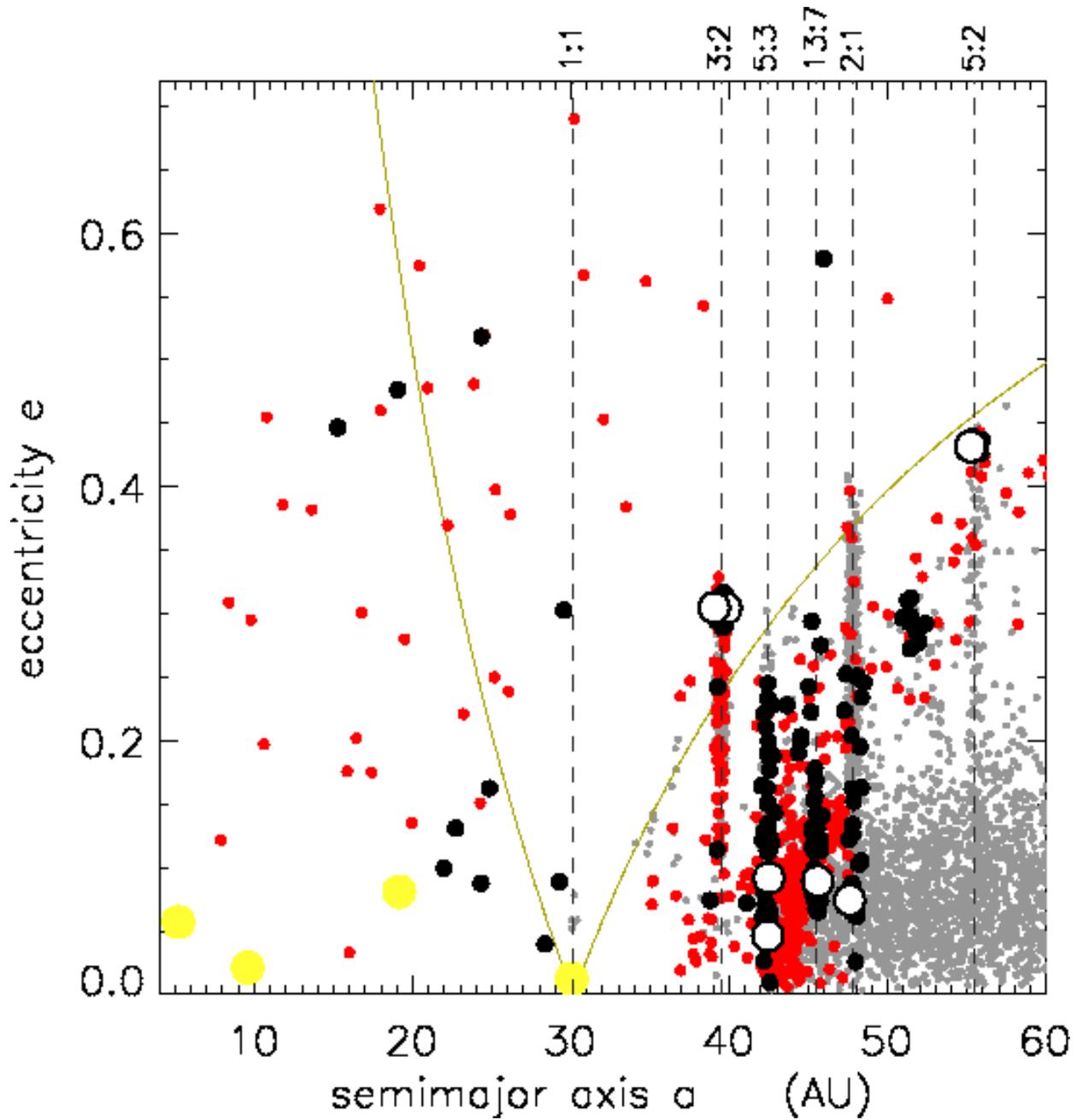}
\figcaption{
  \label{centaur_fig}
  Open circles are the orbits of the simulation's seven Centaurs
  at time $t=10^8$ years, while black dots indicate their subsequent motions.
  Gray dots are the simulation's endstate at time $t=4.5\times10^9$ years,
  red dots indicate multi--opposition KBOs and Centaurs, and yellow dots
  are the giant planets' final orbits. Neptune's mean--motion resonances
  are indicated, and the curve is the threshold for Neptune--crossing orbits.
}
\end{figure}

It is generally accepted that Centaurs are those bodies that have diffused inwards 
from the Kuiper Belt into orbits that cross the giant planets (e.g., \citealt{DQT88}). Presently
there  are 27 known Centaurs observed for more than one opposition; these are the
red dots in Fig.\ \ref{centaur_fig} having $a<30$ AU. Since
planet--crossers are quickly ejected or accreted,
Centaurs have short dynamical lifetimes of only $\sim10^7$ years 
\citep{LD97, TM03}.
Consequently, the density of these `escapees from the Kuiper Belt'
(e.g., \citealt{SC96}) is very tenuous inside of $a=30$ AU 
(see Fig.\ \ref{erosion}). Indeed, only $N_C^\star=7$ Centaurs are detected 
during the final $\Delta T=2$ billion years of our simulation that
was only sparsely time--sampled once every $\Delta t=10^8$ years,
so the instantaneous number of Centaurs is 
$N_C=N_C^\star(\Delta t/\Delta T)=0.35$ at the end of the simulation. 
There are also $N_{MB}=565$ bodies in the Main Belt, so the
Centaur/Main Belt ratio is provisionally estimated
at $r_{C/MB}=6.2\times10^{-4}$.

\placefigure{centaur_fig}

The open circles on Fig.\ \ref{centaur_fig} show the orbital elements
of these seven Centaurs at time $t=10^8$ years, which is at a time when
planet migration has only recently ceased. Thus the open dots
indicate the locations where Neptune has parked these
proto--Centaurs in the Kuiper Belt. 
Note that all seven Centaurs originate from sites in/near
Neptune's mean--motion resonances, namely, the 3:2, 5:3, 13:7, 2:1, and the
5:2. Their subsequent motions at times $t>10^8$ years
are shown as black dots
(again, poorly time--sampled), which show that the eccentricities
of nearly all proto--Centaurs initially wander up--and--down with constant $a$
until they have a close encounter with Neptune, scatter off that planet,
make a brief apparition in the $a<30$ AU Centaur zone, and then are
quickly removed from the system. 

These seven bodies have initial semimajor
axes of $28\le a\le48$ AU at time $t=0$, so Centaurs can also be regarded as
samples that have been drawn from a wide swath of the outer solar nebula.
Fig.\ \ref{centaur_fig} also shows that the simulated Centaurs are concentrated
just inside of Neptune's orbit; their mean heliocentric distance is
$r=26\pm3$ AU, and their mean inclination is $i=16\pm10^\circ$.
Note also that three of the seven Centaurs emerged from the 3:2 and 2:1 resonances,
which Section \ref{census} showed to be heavily depleted relative
to the model's predictions. Consequently, the Centaur/Main Belt
ratio reported above should instead be interpreted as an upper limit, e.g., 
$r_{C/MB}<6.2\times10^{-4}$. It is shown later in Section \ref{calibration}
that our model predicts that 
there are $N_{MB}\simeq1.3\times10^5$ Main Belt KBOs having
radii $R>50$ km, so this model also predicts that there are
$N_C=r_{C/MB}N_{MB}<80$ similarly--sized Centaurs. 

Although the Centaur upper limit reported here is comparable to the 
population that \cite{SJTBA00} infer from the Centaur luminosity function,
there is still a prominent disconnect in the heliocentric distances of the 
simulated and observed populations; our simulated Centaurs 
all reside at $r>22$ AU, while the three
Centaurs that \cite{SJTBA00} used to construct the Centaur luminosity 
were detected at heliocentric distances of $r< 19$ AU. One possible
interpretation of this excess of Centaurs at $r\lesssim20$ AU 
is that Centaurs may be breaking up and spawning 
new Centaurs (e.g., \citealt{PR94}) as they wander among the giant planets.
Finally, we note that deep, wide--angle surveys of the Kuiper
Belt, such as the Legacy Survey that is currently being implemented
at the Canada France Hawaii Telescope, may soon reveal the existence
of the more distant Centaurs that is anticipated by this model to reside at
distances of $23\lesssim r\lesssim29$ AU.

%%%%%%%%%%%%%%%%%%%%%%%%%%%%%%%%%%%%%%%%
% Trojans
%%%%%%%%%%%%%%%%%%%%%%%%%%%%%%%%%%%%%%%%
\section{Neptune's Trojans}
\label{troj}

\begin{figure}
\epsscale{1.0}
\plotone{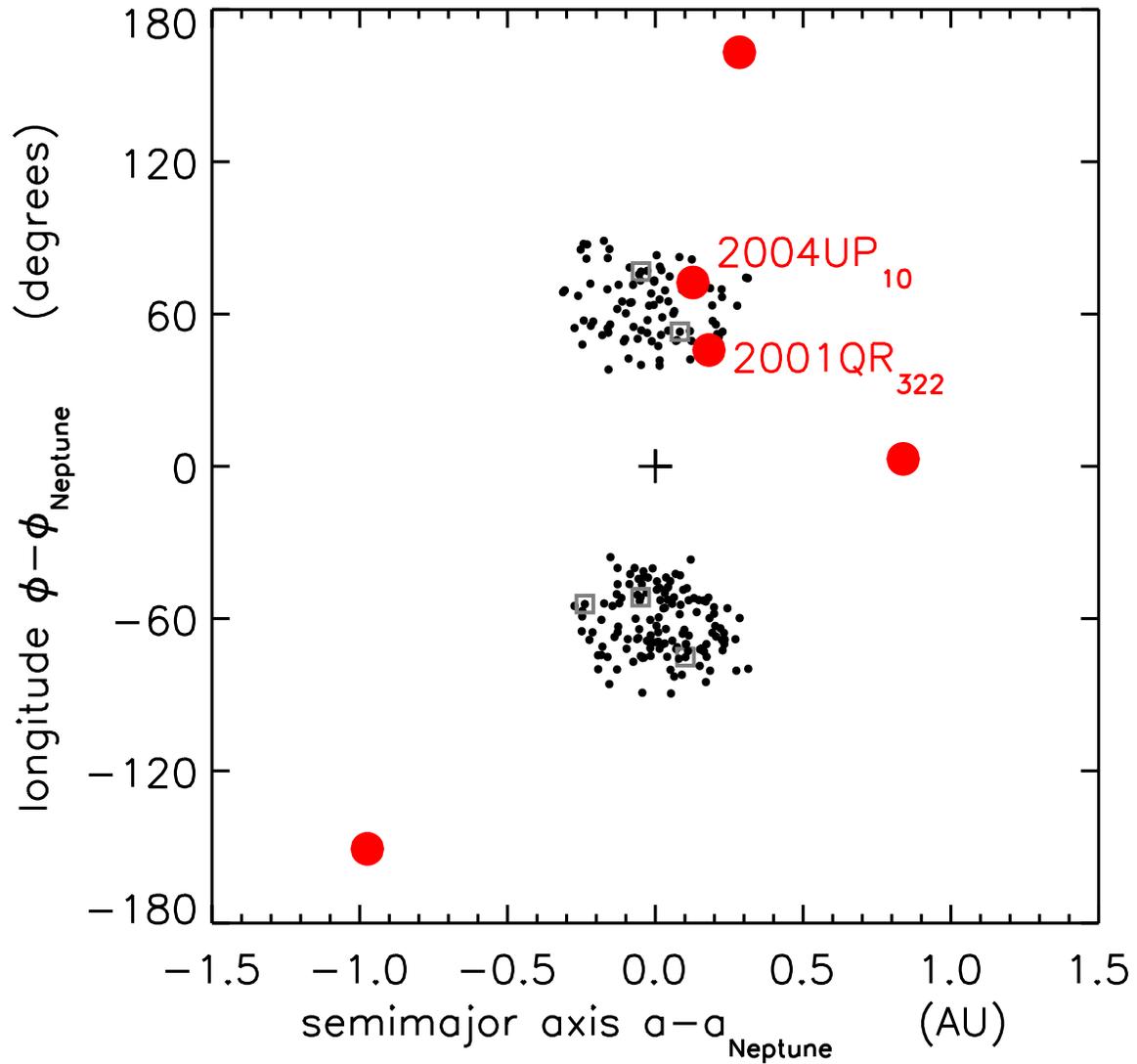}
\figcaption{
  \label{trojans}
  The black dots are the longitudes of five simulated trojans versus their semimajor
  axes, all relative to Neptune's, and sampled every $10^8$ 
  years during the entire simulation. 
  Neptune lies at the $+$. Small squares show the final positions of these trojans,
  indicating that two trojans lie at the leading Lagrange point, with three 
  trailing. Red dots indicate three nearby `field' KBOs
  as well as Neptune's two known Trojans, 2001 QR$_{322}$ and 2004 UP$_{10}$.
}
\end{figure}

Figure \ref{hot belt} also shows that $N_T=5$ particles managed to survive
the length of the simulation at Neptune's 1:1 resonance. These simulated
particles are of course Neptune's Trojans, of which two are presently known:
2001 QR$_{322}$ \citep{Cetal03} and 2004 UP$_{10}$ 
\citep{STJM05}. For this simulation the
Trojan/MB ratio is $r_{T:MB}=N_T/N_{MB}=8.8\times10^{-3}$,
where $N_{MB}=565$ is the number of survivors that persist in the Main Belt.
The spatial coordinates of the two observed and five simulated
Trojans are shown in Figure \ref{trojans},
which indicates that these particles can roam about with longitudes
of $\Delta\phi\simeq\pm30^\circ$
from Neptune's triangular Lagrange points with semimajor axes of
$\Delta a\simeq\pm0.32$ AU from Neptune's. The extent of these Trojan sites are
similar to that seen in integrations by \cite{HW93} and \cite{ND02}.
Note that no special effort was made to start any of the
simulated particles at Neptune's
Lagrange points. Rather, all particles were distributed randomly about a disk
according to a smooth surface density law, with the inner
edge of the disk being well inside of Neptune's initial tadpole
region. In our simulation the five survivors had initial semimajor
axes of $\Delta a\simeq\pm0.28$ AU from Neptune's initial $a$,
and there were a total of 68 particles initially in Neptune's Trojan source region
({\it i.e.}, $|\Delta a|\le0.28$ AU and $|\Delta\phi|\le30^\circ$), so
the surviving Trojan fraction is about $7\%$. This survival fraction is comparable
to that obtained by \cite{KM04} in a similar simulation.
That work also showed that as planets migrate,
several secondary resonances sweep across the 1:1, which result
in a heavy loss of Neptune's Trojans during the migration epoch. 

\placefigure{trojans}

Neptune's Trojans are of interest since they might
place constraints on some models of the early evolution of the outer
Solar System. For example, \cite{Tetal99, Tetal02} postulate that Neptune
originally formed in the vicinity of Jupiter and Saturn and was tossed outwards
after scattering off the larger planets. But the existence of 2001 QR$_{322}$
and 2004 UP$_{10}$ might cast doubt on this scenario since Trojans
seem unlikely to persevere at Neptune's Langrange during such a scattering event.
However it has since been shown
that a recently--scattered Neptune can still acquire its Trojans later
as that planet's orbit is circularized by a dense Kuiper Belt 
(Levison 2005, personal communication). 
It is also conceivable that Neptune may have captured its Trojan from a heliocentric
orbit {\em after} Neptune's orbit has settled down. Although
\cite{KM04} shows that the direct capture of Trojans from heliocentric
space is rare and results in only transient Trojans, 
\cite{CL05} show that mutual collisions may have inserted small bodies into
stable orbits at Neptune's Lagrange points after its orbit has circularized.

%%%%%%%%%%%%%%%%%%%%%%%%%%%%%%%%%%%%%%%%%%%%%%%%%%%%%%%%%%%%%%%%%%%%%%%%%
% The Extended Scattered Disk.      %
%%%%%%%%%%%%%%%%%%%%%%%%%%%%%%%%%%%%%%%%%%%%%%%%%%%%%%%%%%%%%%%%%%%%%%%%%
\section{The Extended Scattered Disk}
\label{extendedSD}

Figure \ref{oss} shows the orbits of those scattered particles that have
been tossed into very wide orbits about the Sun. Most of the simulated
scattered particles have perihelia between $30\lesssim q\lesssim 40$,
as do most of the observed scattered KBOs. However there are two
exceptions to this rule, namely, 2000 CR$_{105}$ and 2003 VB$_{12}$
(also known as Sedna) which have respective perihelia of $q=44.14\pm0.02$
\citep{Getal02} and $q=76\pm4$ AU \citep{Betal04}. \cite{Getal02}
classify those Scattered KBOs having perihelia higher than $q\simeq40$ AU
as members of a so--called extended Scattered Disk. Sedna's large radius of
$R\sim1000$ km makes this object a particular curiosity since its
discovery circumstance suggests that there may be a few hundred
other unseen Senda--sized objects \citep{Betal04}.
Since Sedna has a mass of $\sim10^{-3}$ M$_\oplus$, the implied mass
that might be hidden in the extended Scattered Disk is
a few tenths of an Earth--mass. Thus Sedna by itself may represent an enormous 
reservoir of unseen mass that is comparable to the `conventional'
Kuiper Belt (see Section \ref{calibration}).

\placefigure{oss}

The extended Scattered Disk is also of dynamical interest since, as \cite{Getal02} 
note,  most dynamical models of the Kuiper Belt (including this one)
generally produce Scattered Objects in lower perihelia orbits having
$q\lesssim40$ AU. \cite{Getal02} review a number of scenarios that might 
explain how a KBO might get promoted from a nearly circular orbit into a wide,
eccentric orbit having $q\gtrsim40$ AU; these include: 
({\it i.}) chaotic diffusion of scattered bodies, ({\it ii.}) 
gravitational scattering  by long--gone massive protoplanets, 
({\it iii.}) scattering by an undiscovered distant
planet, and ({\it iv.}) scattering by a single star that passes to within
$\sim100$ AU of the Sun.
However all of these scenarios are problematic. 
For instance, billion--year integrations of $\sim10^4$ particles in chaotic
Neptune--scattered orbits fail to diffuse into orbits having perihelia as
high as that of 2000 CR$_{105}$ \citep{Getal02}.
\cite{Metal02} also cast doubt on scenarios ({\it ii--iii.}) 
by showing that $\sim20\%$ of any distant population of
protoplanets would have persisted over
the age of the Solar System, and that some fraction of these large
bodies should already have been discovered by one of the various wide--angle
Kuiper Belt surveys. Scenario ({\it iv.}) is also in doubt since
simulations of a close encounter with a single star generally produce 
disturbances in an outer Kuiper Belt that is quite unlike that seen
in the observed Belt (e.g., \citealt{Ietal00}). However \cite{F00} have shown
that repeated encounters by more distant stars can produce Sedna--like
orbits. This may have occurred early while the Sun was still a member of
the open cluster from which it presumably formed. In this scenario, the giant planets
scatter small bodies into wide orbits of $a\sim100$ to 1000 AU,
which the nearby cluster stars then perturb into Sedna--like orbits having higher 
perihelia. Scattering by a passing star was recently re--examined
by \cite{ML04}, and their simulations also support this scenario.

Although our simulations did not produce any Sedna--like objects 
in orbits that are well--decoupled from the giant planets, we did find
a single scattered object in a 2000 CR$_{105}$--like orbit in the Extended
Scattered Disk with a semimajor axis of $a=92$ AU and a perihelion of 
$q=43.5$ AU (the large black dot in Figure \ref{oss}). 
The orbital history of this scattered
particle is shown in Fig.\ \ref{p2847}, which shows
that as the particle inhabited Neptune's 16:3 resonance
during times $2\lesssim t\lesssim3$ Gyrs, some process 
raised this Scattered particle's perihelion up and into the Extended
Scattered Disk on a billion--year timescale. This kind of behavior was first
reported in \cite{LD97} and \cite{DL97}, whose simulations
also show that some scattered particles can achieve high perihelia orbits
while in or near mean--motion resonances.

However we have not identified any particular resonance as being responsible
for raising the perihelia of our one CR$_{105}$--candidate shown in
Fig.\ \ref{p2847}. 
For instance, a Kozai resonance is not implicated since the argument of
perihelion $\omega$ does not librate. The possibility of other
Pluto--like `super--resonances' (e.g., \citealt{MW97}) was also examined; this
is the libration of a resonance angle of the form
$\phi=j(\omega-\omega_N)-k(\Omega-\Omega_N)$, where $\Omega$
is the particle's longitude of the ascending node and the $N$
subscript refers to Neptune's orbit elements. Angles having 
$0\le|j|,|k|\le 10$ were examined, and although
the angle $\phi=5(\Omega-\Omega_N)$ angle did in fact librate for
about 1 Gyrs, that occurred well after the
time when the particle's $q$ was raised.
A resonance involving interactions with multiple planets is also unlikely
since the particle's Tisserand parameter $T$ (which is simply its Jacobi integral 
sans the interaction energy due to Neptune) was well--preserved during these times.
Although the particular
mechanism that drove this particle into the Extended
Scattered Disk in not understood, this particle does demonstrate that it is
indeed possible for Scattered particles to diffuse into the Extended Scattered 
Disk via planetary perturbations alone, with other external agents 
(like a stellar encounters) being absent. This transport from the 
Scattered Disk to the Extended Scattered Disk via mean--motion resonances
is also evident in the simulations of \cite{LD97} and \cite{DL97}.
However this transport has an extremely low flux since only
one of the $\sim5\times10^3$ particles initially in the Scattered Disk
did manage to enter the Extended
Scattered Disk and persist over the age of the Solar System. External perturbations
from passing stars \citep{F00, ML04} may indeed be more
effective at producing members of the Extended Scattered Disk.

\placefigure{p2847}

Note also the simulated particles represented by crosses in Fig.\ \ref{oss}.
Even though their perihelia of $42<q<54$ AU might suggest that they also
inhabit the domain of the Extended Scattered Disk, they are in fact resonant
particles that were trapped at the 3:1, 7:2, and 4:1 resonances during the migration
epoch. Most of these particles
have libration amplitudes less than $|\Delta\phi_{jk}|\lesssim90^\circ$.
These particles also had initial semimajor axes of ${a>47}$ AU, which is
noteworthy since, if any resonant KBOs are ever discovered in these orbits,
they could be interpreted as evidence that the outer edge of the Solar System lies
beyond ${a>47}$ AU. However that interpretation would still be ambiguous, 
since Neptune--scattered evaders, which originated from smaller semimajor axes,
can also settle into these same resonances---see Fig.\ 18 of \cite{G03} 
for an example.

%%%%%%%%%%%%%%%%%%%%%%%%%%%%%%%%%%%%%%%%%%%%%%%%%%%%%%%%%%%%%%%%%%%%%%%%%
% Scattered Disk.       %
%%%%%%%%%%%%%%%%%%%%%%%%%%%%%%%%%%%%%%%%%%%%%%%%%%%%%%%%%%%%%%%%%%%%%%%%%
\subsection{The Scattered Disk}
\label{SD}

\begin{figure}
\epsscale{1.0}
\plotone{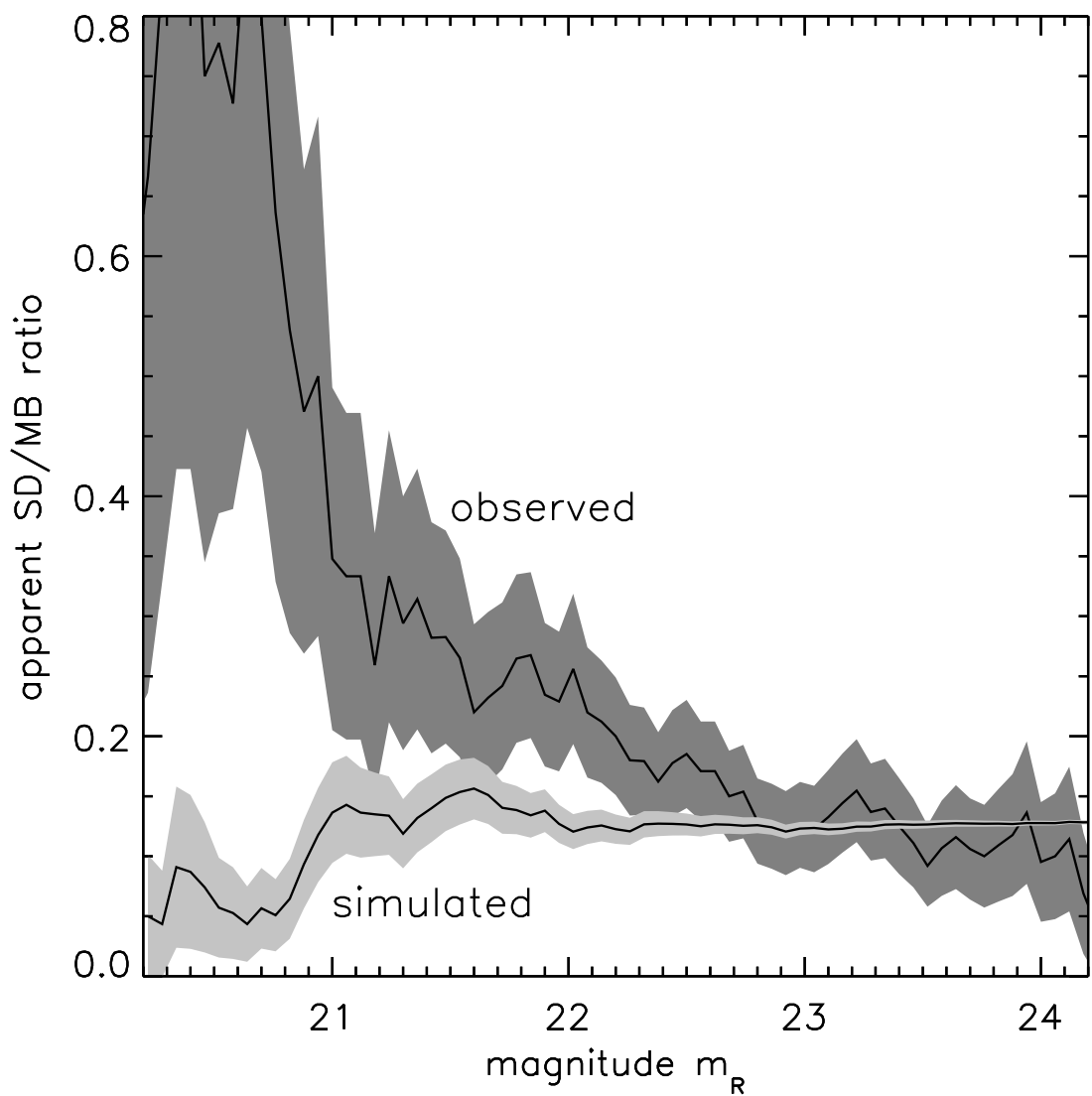}
\figcaption{
  \label{sd_ratio}
  The dark gray curve is the observed ratio of Scattered Disk (SD) objects
  in orbits having $50<a<150$ AU and perihelia $28<q<40$ AU, relative to Main Belt
  (MB) objects, versus apparent magnitude $m_R$. The light gray curve is the 
  simulated ratio.
}
\end{figure}

Figure \ref{sd_ratio} shows the apparent abundance of so--called Scattered Disk
objects relative to the Main Belt, as predicted by the Nbody/Monte Carlo model.
There are $N_{SD}=1.9\times10^4$ Monte Carlo bodies in orbits having 
$50<a<150$ AU and perihelia $28<q<40$ AU, while 
$N_{MB}=1.0\times10^5$ Monte Carlo particles survive in the Main Belt, so the
model predicts an intrinsic SD/MB ratio of
$r_{SD/MB}=N_{SD}/N_{MB}=0.19$.
The apparent ratio of these two populations is $\sim0.1$, also shown in 
Fig.\ \ref{sd_ratio}. Note that the intrinsic SD/MB ratio inferred here is
about one--fourth that reported in \cite{TJL01}. 

\placefigure{sd_ratio}

%%%%%%%%%%%%%%%%%%%%%%%%%%%%%%%%%%%%%%%%%%%%%%%%%%%%%%%%%%%%%%%%%%%%%%%%%
% Calibrate.        %
%%%%%%%%%%%%%%%%%%%%%%%%%%%%%%%%%%%%%%%%%%%%%%%%%%%%%%%%%%%%%%%%%%%%%%%%%
\section{Calibration}
\label{calibration}

\begin{figure}
\epsscale{1.0}
\plotone{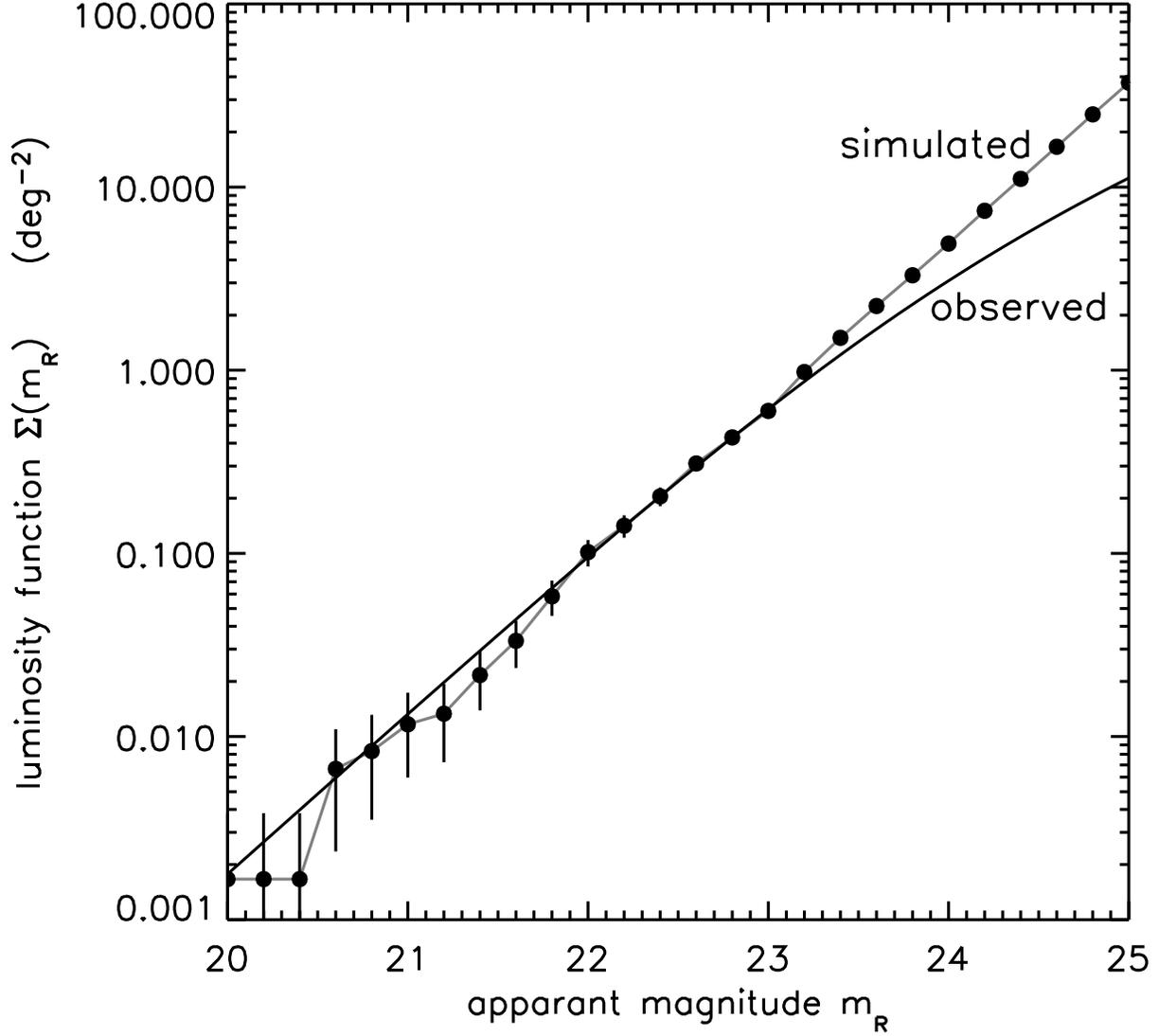}
\figcaption{
  \label{lum_fn_fit}
  The smooth black curve is the Kuiper Belt's observed ecliptic luminosity function 
  $\Sigma(m_R)$, obtained by integrating the differential luminosity
  function $d\Sigma(m_R)/dm_R$ reported in \cite{Bernstein04}; note that it
  breaks to a shallower slope at magnitudes $m_{break}\simeq24$.
  Dots give the simulation's ecliptic luminosity function for particles in a 
  truncated Kuiper Belt (see Section \ref{calibration}) 
  having latitudes within $0.5^\circ$ of the ecliptic.
  Error bars are for Poisson counting statistics.
}
\end{figure}

The ecliptic luminosity function of \cite{Bernstein04} is shown in Fig.\
\ref{lum_fn_fit}, and its bright end varies as
$\Sigma(m_R)=10^{\alpha(m-m_0)}$ deg$^{-2}$ where $\alpha=0.88$ and $m_0=23.1$,
for magnitudes $m_R<m_{break}=24$.
This luminosity function gives the number density of KBOs near the ecliptic that
are brighter than magnitude $m_R$. Since this curve scales with the total number of
KBOs, it can be used to calibrate
the simulation to determine the total number of objects in the Kuiper Belt.

\placefigure{lum_fn_fit}

Sections \ref{census}--\ref{edge} show that the observed 3:2 and 2:1 populations are
severely depleted relative to model predictions, and that bodies in an Outer Belt 
beyond the 2:1 are either absent or too faint to be seen. To account for these 
depletions, the truncated Kuiper Belt similar to that of Section \ref{edge} is adopted; 
this Belt is formed by discarding any bodies orbiting beyond the 2:1, as well 
as all bodies orbiting within $\Delta a=0.6$ AU of Neptune's 3:2 and 2:1 
resonances. 
There are $N_{Nbody}=587$ Nbody particles in this truncated Kuiper Belt,
and they are replicated $N_{reps}=10^4$ times with
sizes and magnitudes assigned to them according to the Monte Carlo method of 
Section \ref{census}, with $Q=4.4$ and $R_{min}=20$ km. 

The simulation's median inclination is low (e.g., Section \ref{i}), 
only $\bar{i}_{sim}=2.7^\circ$, which is much lower than the median inclination 
$\bar{i}_{obs}=15.6^\circ$ that is inferred from the debiased KBO inclination 
distribution reported by \cite{B01}.
Due to these low inclinations, the simulation's ecliptic luminosity function
would thus be artificially overdense by a factor 
$f_i=\bar{i}_{obs}/\bar{i}_{sim}\simeq5.8$, so it is revised
downwards by this factor to compensate.
The simulation's $\Sigma(m_R)$  is then multiplied by a factor 
$f_N=3.5$ to fit it to the bright end of the observed luminosity function; this
accounts for the different populations in the simulated and observed
Kuiper Belts, and results in the curve shown in Fig.\ \ref{lum_fn_fit}.
The size distribution adopted here is valid down to a radius of about
$R_{break}=65$ km (see Section \ref{census}), so the inferred number of KBOs larger
than $R_{break}$ is 
$N_{break}=N_{Nbody}N_{reps}f_N(R_{break}/R_{min})^{-Q}\simeq1.1\times10^5$.
To estimate the total number of KBOs larger than the fiducial radius of
$R_{50}=50$ km, note that the faint end of the observed luminosity function
has a logarithmic slope of $\alpha_{faint}=0.32$ \citep{Bernstein04}, 
which implies a power--law index of $Q_{faint}=5\alpha_{faint}=1.6$ for bodies 
having radii $R<R_{break}$. The total number of bodies larger than $R_{50}$ is
thus $N_{50}=N_{break}(R_{50}/R_{break})^{-Q_{faint}}\simeq1.7\times10^5$.

The total mass of these bodies is obtained from their 
cumulative size distribution, which for the large bodies with $R>R_{break}$
can be written $N(R)=N_{break}(R/R_{break})^{-Q}$.
The differential size distribution is then
$dN(R)=|dN/dR|dR$, and if $M(R)=$ mass of a body having a radius $R$,
the total mass of bodies having radii in the interval $R_{break}<R<R_{max}$ is
\begin{eqnarray}
  \label{Mtotal}
  M(R>R_{break})&=&\int_{R_{break}}^{R_{max}}M(R')dN(R')=
    \frac{Q}{Q-3}\left[1-\left(\frac{R_{break}}{R_{max}}\right)^{Q-3}\right]
    N_{break}M_{break}
\end{eqnarray}
where $M_{break}=M(R_{break})\simeq1.2\times10^{21}
(\rho/\mbox{1 gm/cm$^3$})(p/0.04)^{-3/2}$ gm, which is 
the mass of a body of radius $R_{break}=65$ km assuming it has a density $\rho$ 
and albedo $p$.
The total mass of KBOs larger than $R_{break}$ with
semimajor axes inside of Neptune's 2:1 is thus
$M(R>R_{break})\simeq0.07(\rho/\mbox{1 gm/cm$^3$})(p/0.04)^{-3/2}$ M$_\oplus$ 
for a $Q=4.4$ size distribution that extends to radii as large as $R_{max}=1000$ km.
To get the total mass of bodies at the fiducial size $R=R_{50}$, add to the above
the mass of bodies in the size interval $R_{50}<R<R_{break}$, which is roughly
$\Delta M\simeq(N_{50}-N_{break})M_{break}$. The total mass of bodies
larger than $R_{50}=50$ km is then $M_{total}=M(R>R_{break})+\Delta M\simeq
0.08(\rho/\mbox{1 gm/cm$^3$})(p/0.04)^{-3/2}$ M$_\oplus$. Note that the
$0.08$ M$_\oplus$ prefactor is a consequence of adopting the oft--employed
Halley albedo of $p=0.04$.  However recent observations 
indicate KBOs have an average albedo of $p\simeq0.1$
\citep{ABM04, GNS05}, which in turn lowers the Kuiper Belt mass to 
$M_{total}\sim0.02$ M$_\oplus$ assuming they have a unit density.

This population estimate is comparable to, but a bit higher than,
previous estimates that rely on far simpler models of the Kuiper Belt. 
For instance, \cite{TJL01} report a Main Belt population of 
$3.8\times10^4$ objects of mass $0.03$ M$_\oplus$
among bodies having radii $R>50$ km. They also estimate the Belt's total 
population to be $1.9\times$ the Main Belt population,
so a total population of $N_{50}\sim7.2\times10^4$ bodies
larger than $R_{50}$ having mass $M_{total}\simeq0.06$ M$_\oplus$ is 
inferred. A similar estimate is also inferred from the HST survey by
\cite{Bernstein04}; according to their Fig.\ 8,
the sky--plane number density of KBOs larger than $R_{50}$
is $\Sigma(R>R_{50})\simeq13$ deg$^{-2}$. Since the Kuiper Belt subtends a total
solid angle of $\Delta\Omega_{total}\simeq8100$ deg$^2$ \citep{B01}, the total number of
KBOs larger than $R_{50}$ is
$N_{50}=\Sigma(R>R_{50})\Delta\Omega_{total}\simeq1.1\times10^5$ having
a total mass of $M_{total}\simeq0.05$ M$_\oplus$.

Sections \ref{census}--\ref{extendedSD} show that the simulated Belt's
various dynamical classes have abundances of $r_{2:1/MB}=0.041$,
$r_{3:2/MB}=0.021$, $r_{C/MB}<6.2\times10^{-4}$, 
$r_{T/MB}<8.8\times10^{-3}$, and $r_{SD/MB}=0.19$
relative to the Main Belt, so the Main Belt fraction is 
$f_{MB}=1-\sum_ir_{i/MB}\simeq0.74$ and thus
there are are $N_{MB}(R>50\mbox{ km})=f_{MB}N_{50}\simeq1.3\times10^{5}$
Main Belt KBOs having radii $R>50$ km. The numerical abundance
of the $i^{th}$ dynamical class is 
$N_i(R>50\mbox{ km})=r_{i/MB}f_{MB}N_{50}$, and its mass is
$M_i(R>50\mbox{ km})=r_{i/MB}f_{MB}M_{total}$ where $M_{total}=0.08$ M$_\oplus$
assuming $\rho=1$ gm/cm$^3$ and $p=0.04$;
these abundances and masses are listed in Table \ref{pop_table}.
The exception is the 3:2 mass estimate which adopts the $Q=2.7$ power--law size 
distribution described in Section \ref{census}; if this subgroup
really does have such a flat size distribution, 
then Eqn.\ (\ref{Mtotal}) must be used to calculate its 
mass\footnote{with the quantities $R_{break}$, $N_{break}$, and $M_{break}$
replaced by $R_{50}$, $r_{3:2/MB}N_{MB}$, and $M(R=50\mbox{ km})$.}.

Note also that the preceding ratios assume that the Main Belt
terminates just inwards of the 2:1 resonance at $a=47.2$ AU. 
If, however, one wishes to adopt an
outer edge at $a=45$ AU, then Section \ref{census} shows that this
reduces the MB population by $40\%$,
so that the ratios $r$ quoted above should then be raised by a factor of
1.7. The exception to this rule are the bodies at the 2:1
resonance---their abundance relative to the MB is unchanged. However
the {\em total} number of KBOs reported here is still insensitive to the detailed 
location of the Main Belt's outer edge, since that number is
obtained by fitting the simulated KBOs' luminosity function to the observed
$\Sigma(m_R)$, which is quite insensitive to the detailed location of the 
Main Belt's outer edge.

It should also be noted that this study employed an initial $\sigma(a)\propto a^{-2}$
disk surface density, but our findings are readily adapted
for an alternate surface density law. For instance, if the canonical 
$\sigma(a)\propto a^{-1.5}$ law were instead desired, then this shallower power law would
result fewer objects trapped in the 3:2 resonance relative to the Main Belt population.
Since the 3:2 objects
are drawn from the $a_{3:2}\sim32$ AU part of the disk, while the Main Belt
objects formed at $a_{MB}\sim44$ AU, this revised surface density
law would reduce the 3:2/MB ratio reported here by a factor
$(a_{3:2}/a_{MB})^{0.5}\sim0.85$, which is a $15\%$ change in relative
abundance. Of course, the 2:1/MB ratio would remain unchanged since their 
source populations are the same. 

Also, we conservatively interpret the abundance of Neptune's Trojans
reported Table \ref{pop_table} as an upper limit on their real abundance.
It was argued in Section \ref{census} that other unmodeled processes,
possibly the scattering of planetesimals by Neptune or amongst themselves,
reduced the trapping efficiency of the 3:2 and 2:1 resonances by factors
of $\sim10$. Thus it is possible that the same unmodeled phenomena might also
have destabilized orbits at Neptune's 1:1 resonance, so the actual number of
Trojan survivors may be smaller than that reported in Section \ref{trojans}.

Finally, upper limits on the abundance of KBOs inhabiting
a hypothetical Outer Belt are reported for the
$50<a<80$ AU zone assuming these bodies have the
shallowest possible size distribution, namely $Q=6.0$ down to $R_{min}=20$ km
(see Section \ref{edge}). In this case, there are at most 
$N_{OB}=1.3\times10^4$ bodies in the OB having radii
of $50<R<250$ km and a total mass of $M_{OB}\simeq0.008$ M$_\oplus$
assuming a density of $\rho=1$ gm/cm$^3$ and an albedo $p=0.04$.

%%%%%%%%%%%%%%%%%%%%%%%%%%%%%%%%%%%%%%%%%%%%%%%%%%%%%%%%%%%%%%%%%%%%%%%%%
% other unmodeled effects.      %
%%%%%%%%%%%%%%%%%%%%%%%%%%%%%%%%%%%%%%%%%%%%%%%%%%%%%%%%%%%%%%%%%%%%%%%%%
\section{Effects not modeled}
\label{unmodeled}

It should be noted that the model used here only
accounts for the Belt's dynamical erosion that is a consequence of Neptune's
gravitational perturbations---it does not account for the collisional
erosion of the Kuiper Belt that is often invoked to account for the Belt's
depleted appearance (e.g., \citealt{S96, KL99}). In particular,
models of KBO accretion, as well as the self--consistent Nbody
simulations of Neptune's migration, all suggest 
that the Kuiper Belt's primordial mass was of order $\sim30$ M$_\oplus$
(\citealt{S96, KL99, HM99, GML04}), which is at least
$\sim400$ times more
than the current mass. However the model used here, which only accounts for the
dynamical erosion, results in a depletion by a factor of about 3 in the 
$30<a<48$ AU zone of Fig.\ \ref{hot belt}. This suggests
that collisional erosion, which is not modeled here, may have been 
responsible for reducing the Belt's mass by an additional 
factor\footnote{Only a `conventional' Kuiper Belt model, like the
one explored here, need invoke additional erosion to reduce
the Kuiper Belt mass by another factor of $\sim100$. 
This is distinct from the
push--out model which need not rely on any collisional
depletion of the Kuiper Belt \citep{LM03}.}
of $\sim100$. Nonetheless, the 
abundance and mass estimates obtained here
should still be reliable provided the Belt's collisional erosion
was relatively uniform across the observable $35\lesssim a\lesssim 50$ AU
zone. If, however, collisional erosion was more vigorous in some parts
of the Belt, and less so in other parts, then the estimates obtained
above will only be accurate in an order--of--magnitude sense. 

A comparable problem also occurs with the model's inclinations. 
Section \ref{calibration} shows that the simulated Kuiper Belt is too thin by a 
factor of $f_i\simeq6$. This is compensated for by reducing the simulation's
luminosity function $\Sigma(m_R)$  by the factor $f_i$, which is equivalent to 
increasing each particles' inclination by this
factor. Again, this crude treatment should still
yield a reliable estimate of the KBO population provided the factor
$f_i$ is uniform everywhere and independent 
of semimajor axis $a$. If, however, $f_i$ is {\em not} independent of $a$,
then this will result in errors in the relative abundances of the
Belt's various subpopulations reported in Table \ref{pop_table}.

We also note that the relative abundances of the Belt's various subpopulations 
are determined by a model that invokes a smooth outward migration by Neptune
by $\Delta a_N\simeq9$ AU, with the results reported in Fig.\ \ref{abundance}. 
That Figure shows that the smooth migration scenario
predicts a combined 3:2 + 2:1 population that is comparable to the 
Main Belt population. This is because
smooth migration is very efficient at trapping particles
at Neptune's resonances, and this results in densely populated resonances.
However a detailed comparison of the model predictions to the
observed abundances indicates that the resonant KBO
population is really only about $5\%$ of the Main Belt population
(see Section \ref{census} and Table \ref{pop_table}). The seemingly low
abundance of resonant KBOs is likely due to unmodeled effects that
may have occurred during the migration epoch, possibly due to 
the mutual scattering that might occur among bodies trapped at resonance,
or perhaps due to the gravitational scattering of large planetesimals by 
Neptune (e.g., \citealt{ZSZZV02, TM04}). 

There is also evidence indicating that a wide swath of the early
Kuiper Belt was stirred up prior to the onset of Neptune's migration.
Recall that simulations of Neptune's outward migration into a dynamically
cold Kuiper Belt is unable to account for the eccentricities of
$e\sim0.1$ observed among Main Belt KBOs (Section \ref{cold belt}).
This suggests that the Belt was stirred up, either prior to or after the onset
of Neptune's migration. However Section \ref{hot belt} shows that
this stirring event likely occurred prior to migration:
migration into a stirred--up Kuiper Belt facilitates trapping at a multitude
of weak, high--order mean motion resonances which, as \cite{Cetal03}
point out, is consistent with the detection of seven KBOs now
known to librate at Neptune's 5:2 resonance.

It is then natural to ask what mechanism might be responsible for 
stirring--up a broad swath of the Kuiper Belt, particularly since
accretion models tell us that KBOs must have formed in a 
dynamically cold environment, {\it i.e.}, the
particles' initial $e$'s and $i$'s were $\lesssim10^{-3}$ (\citealt{S96, KL99}).
Note that this disturbance was probably not due to gravitational
stirring by a number of long--gone protoplanets since, as \cite{Metal02}
point out, a sizable fraction of such bodies would still persist in the 
Kuiper Belt and would likely have been discovered by now.

Note that this stirring mechanism must also have had a large reach since
it must afflict KBOs across the entire Main Belt,
at least out to Neptune's 2:1 resonance. One mechanism that comes to mind is
secular resonance sweeping, which 
is the only mechanism known to us that 
might stir eccentricities in the Belt up to $e\sim0.1$ across its entire 
width \citep{NI00}. However this $e$--excitation is coherent in the sense
that neighboring particles will have similar eccentricities. So it is
unclear whether secular resonance sweeping of the Main Belt, 
which would then followed by sweeping
mean--motion resonances due to Neptune's migration, will 
result in the range of eccentricities that is seen in Fig.\ \ref{hot belt}.
Secular resonance sweeping
is a consequence of the dispersal of the solar nebula gas; the removal
of that gas alters the giant planets' precession rates which in turn
shifts the location of secular resonances \citep{W81}. 
The magnitude of the disturbance caused by a sweeping secular resonance
depends on the timescale over which the nebula is depleted; a
longer depletion timescales $\tau_{dep}$ 
results in larger eccentricity--pumping. The simulations of nebula dispersal
by \cite{NI00} show that a disturbance of $e\sim0.1$ across much of the Kuiper
Belt requires a nebula depletion timescale of 
$10^6\lesssim\tau_{dep}\lesssim10^7$ years (but see also \citealt{HW02}).

%%%%%%%%%%%%%%%%%%%%%%%%%%%%%%%%%%%%%%%%%%%%%%%%%%%%%%%%%%%%%%%%%%%%%%%%%
% Discussion.        %
%%%%%%%%%%%%%%%%%%%%%%%%%%%%%%%%%%%%%%%%%%%%%%%%%%%%%%%%%%%%%%%%%%%%%%%%%
\section{Discussion}
\label{discussion}

One of the goals of this study is to determine how the adoption
of a particular Kuiper Belt model might affect an
assessment of the Belt's total
population and mass. Note that some
models of the Belt assume that the KBOs are distributed according
to a primordial surface density distribution that might vary with distance
at $\sigma(r)\propto r^{-2}$ or so (e.g., \citealt{JL95, TJL01}), while the 
KBOs in another model are essentially equidistant \citep{Bernstein04}.
However Fig.\ \ref{erosion} shows that a realistic Kuiper Belt has been
eroded from the inside--out by the giant planets' gravitational perturbations, 
which suggests that the earlier models might
not apply. However it turns out that an estimate of the total KBO
population does {\em not} depend strongly upon a particular
model's radial variation. As Section \ref{calibration} shows,
all three models yield population
estimates that are within a factor of $\sim2$
of each other. This is because the observable KBOs really do inhabit 
a relatively narrow Belt centered on $r\simeq45$ AU
having a radial half--width that is only $\Delta r\sim4$ AU
(see Fig.\ \ref{erosion}), so the the assumption of equidistant KBOs
(e.g., \citealt{Bernstein04}) appears to be good enough. 

It should also 
be noted that the magnitude interval over which a model Kuiper Belt
can be compared to the observed Belt is given by the brightness
of those KBOs having reliable orbits, and this sample is presently dominated 
by bodies having a relatively limited magnitude range
of only $21\lesssim m_R\lesssim 24$. Further testing of this model, as
well as the development of alternative models of the Belt,
would be greatly facilitated if they could be compared to a larger
sample of multi--opposition KBOs having reliable orbits and also exhibiting a
broader range of apparent magnitudes and sizes. 
This larger KBO sample would be very 
useful in many ways. For example, it could be used to test the 
possibility that the various Kuiper Belt subpopulations do exhibit
variations in their size distributions (e.g., Section \ref{census}).
This larger sample
might also permit a better understanding of certain rare and unusual
KBOs, such as those that inhabit the Extended Scattered Disk 
(Section \ref{extendedSD}).
A deeper understanding of the Kuiper Belt,
and what the Belt tells us about the early evolution of the outer Solar System,
would be facilitated by deeper KBO surveys over larger portions of the sky in 
a systematic way that leads to efficient KBO recoveries and reliable
orbit determinations.

%%%%%%%%%%%%%%%%%%%%%%%%%%%%%%%%%%%%%%%%%%%%%%%%%%%%%%%%%%%%%%%%%%%%%%%%%
% Summary.        %
%%%%%%%%%%%%%%%%%%%%%%%%%%%%%%%%%%%%%%%%%%%%%%%%%%%%%%%%%%%%%%%%%%%%%%%%%
\subsection{summary of findings}
\label{summary}

\begin{itemize}

\item Accretion models have shown that Kuiper Belt Objects
must have formed in a dynamically cold environment where the initial KBO seeds
had nearly circular and coplanar orbits with eccentricities and inclinations
$\lesssim0.001$ \citep{S96, KL99}. Simulations 
of Neptune's outwards migration into a dynamically
cold Kuiper Belt, described in Section \ref{cold}, 
show that the survivors in the Main Belt still maintain
low eccentricities and inclinations. However this 
conflicts with the Main Belt's observed $e$'s
and $\sin i$'s of $\sim0.1$. 
This discrepancy suggests that some other process has
also stirred--up the Kuiper Belt. This stirring event could have occurred prior
to or after the onset of planet migration.

\item The existence of several KBOs librating at Neptune's
5:2 resonance suggest that this stirring event occurred prior to the
onset of planet migration.
Simulations by \cite{Cetal03} have shown that if Neptune migrates
into a stirred--up Kuiper Belt having eccentricities of $e\sim0.1$,
then trapping at Neptune's higher--order resonances, such as the 5:2, 
becomes more efficient. This result is confirmed by a higher--resolution
study of this phenomena described in Section \ref{hot}, which reveals 
that additional trapping also
occurs at a number of exotic mean motion resonances
like the 11:6, 13:7, 13:6, 9:4, 7:3, 12:5, 8:3, 3:1, 7:2, and the 4:1;
such resonances are not populated when Neptune migrates 
into a dynamically cold disk. Not surprisingly, 
Neptune's migration into a previously
stirred--up Kuiper Belt also accounts for the eccentricities of $e\sim0.1$
observed in the Main Belt.

\item However the planet--migration scenario investigated here does not account
for the observed KBOs having inclinations above $i\sim15^\circ$ 
(Section \ref{i}), which is the main deficiency of this model. This
is a serious discrepancy since half of all KBOs have inclinations $i>15^\circ$
according to the debiased inclination distribution reported by \cite{B01}.

\item Neptune's migration into a stirred--up Kuiper Belt traps
particles in eccentric orbits at a number of resonances beyond $a=50$ AU,
the most prominent of these being the 5:2 and the 3:1. Many of these
distant particles that are trapped at semimajor axes $a>50$ AU also
have perihelia $30\lesssim q\lesssim 40$ AU, which is the domain
conventionally known as the Scattered Disk. However Section
\ref{hot} shows that only about 10\% of the simulated particles that inhabit
the so--called Scattered Disk or the Extended Scattered Disk
(such as the gray zone in Fig. \ref{sd} where $50<a<80$ AU and $e>0.25$)
are truly scattered particles.
The vast majority of these particles never had a close encounter with Neptune;
rather, they were placed in these wide, eccentric orbits
by Neptune's sweeping mean--motion resonances. Note that the origin 
of these bodies as being due to resonant trapping
is very distinct from the scattering scenario originally
suggested by \cite{DL97}.

\item Of the $10^4$ particles simulated here, only one managed to persist
over the age of the Solar System in the Extended
Scattered Disk, which is loosely defined as scattered orbits
having perihelia $q\gtrsim40$ AU. This particle's 
orbit is qualitatively similar to 2000 CR$_{105}$ which has a 
perihelion of $q=44$ AU. However our simulations did not produce any extreme 
members of the Extended Scattered Disk that are similar to
2003 VB$_{12}$ (Sedna), which has a perihelion of $q=76$ AU.

\item The output of the Nbody model is 
coupled to a Monte Carlo model that assigns
radii $R$ to the simulated particles according to a power--law type
cumulative size distribution that varies as $N(R)\propto R^{-Q}$. 
Magnitudes are computed for the simulation's particles, which then allows us
to directly compare the simulated Belt to the observed Belt in a manner
that accounts for telescopic selection effects. Section \ref{census} 
compares the observed abundance of 2:1 objects to known Main Belt objects, and
it is shown that the observed 2:1 population is underdense by a factor of 20 
relative to model predictions. Similarly, the observed
3:2 population is also depleted relative to model
expectations. Another curious feature of the 3:2 is its lower than expected
abundance (relative to the Main Belt KBOs) of fainter KBOs having magnitudes 
$m_R\gtrsim23.5$. Section \ref{census} shows that this dearth of fainter KBOs at
the 3:2 can be interpreted as a dearth of small bodies,
which implies that the 3:2 population has a $Q\simeq2.7$
size distribution that is substantially shallower than the canonical $Q=4.4$ 
power--law that holds for the larger members of the Main Belt.

\item The simulated Centaurs are quite sparse owing to their short dynamical 
lifetimes; only seven Centaurs were detected in the simulation during
its final 2 Gyrs. Interestingly, all seven originated at/near Neptune's
mean motion resonances in the Kuiper Belt (Section \ref{centaurs}).
The model puts an upper limit of $N_C\lesssim80$ Centaurs having radii
larger than $R=50$ km, assuming they have a material density 
$\rho=1$ gm/cm$^3$ and an albedo $p=0.04$. It should be noted that all of the
simulated Centaurs inhabit heliocentric distances of $r>22$ AU,
while the three Centaurs reported in \cite{SJTBA00} were detected
at $r<19$ AU. If the simulated Centaurs are representative of reality, 
then this discrepancy in their heliocentric distances may
indicate that Centaurs can break up and spawn additional Centaurs
(e.g., \citealt{PR94}) after evolving inwards from the Kuiper Belt.

\item This model also estimates that there are at most
$N_C\sim1100$ Trojans larger than $R=50$ km 
having a total mass of $M_C\sim5\times10^{-4}$ M$_\oplus$
orbiting at Neptune's triangular  Lagrange points, 
assuming the usual $\rho=1$ gm/cm$^3$ and $p=0.04$.

\item The absence of any distant KBOs having low eccentricities at $a>50$ AU 
places tight upper limits on the abundance
of any KBOs that might inhabit a hypothetical Outer Belt. Several upper
limits are inferred from this null result: ({\it i.}) the primordial
density of Outer Belt objects beyond 50 AU is smaller than the primordial
Main Belt density by a factor $f\gtrsim100$, ({\it ii.}) these distant
KBOs are fainter than $m^\star\simeq24.5$ and thus have radii
smaller than $R\simeq80(p/0.04)^{-1/2}$ km, 
({\it iii.}) the cumulative size distribution of Outer Belt objects is steep, 
having a power--law index of $Q>6.0$, or ({\it iv.}) some combination of the
above.

\item The luminosity function of the Nbody/Monte Carlo model is fitted
to the KBO's observed luminosity function, which then yields an estimate
of the Belt's total population of $N\sim1.7\times10^5$ KBOs larger than
$R\simeq50(p/0.04)^{-1/2}$ km having a total mass of
$M_{total}\sim0.08(\rho/\mbox{1 gm/cm$^3$})(p/0.04)^{-3/2}$ M$_\oplus$.
The population and mass of the Belt's various subclasses (e.g., 
Centaurs, Neptune Trojans, 3:2 and 2:1 populations, the Main Belt,
and the Scattered Disk) are also assessed in Section \ref{calibration}
and listed in Table \ref{pop_table}.

\end{itemize}

%%%%%%%%%%%%%%%%%%%%%%%%%%%%%%%%%%%%%%%%%%%%%%%%%%%%%%%%%%%%%%%%%%%%%%%%%%
% Acknowledgements       %
%%%%%%%%%%%%%%%%%%%%%%%%%%%%%%%%%%%%%%%%%%%%%%%%%%%%%%%%%%%%%%%%%%%%%%%%%
\newpage
\acknowledgements

\begin{center}
  {\bf Acknowledgements}
\end{center}

The authors thank Matt Holman for his comments on this paper,
Rodney Gomes for discussions of the Neptune--evader
mechanism, and Hal Levison for his review this paper.
The derivation of Brouwer's $B$ integral appearing in Appendix 
\ref{appendix_BI} was suggested to us several years ago by Bill Ward.
Support for this research was provided to JMH by a Discovery Grant
from the Natural Sciences and Engineering Research Council of Canada
(NSERC). Several of the simulations reported here were executed on the
Institute for Computational Astrophysics (ICA) Pluto computer cluster,
which is funded by a grant from the Canada Foundation for Innovation (CFI).
Additional simulations were also performed on the McKenzie computer
cluster at the 
Canadian Institute for Theoretical Astrophysics (CITA) at the University
of Toronto; those machines are funded by the CFI as well as the Ontario 
Innovation Trust (OIT). An early generation of these simulations was also
performed while JMH was in residence at the Lunar and Planetary
Institute (LPI), and that portion of this research was supported by the
National Aeronautics and Space Administration (NASA) via an
Origins of Solar Systems grant NAG5-10946.
RM acknowledges research support from NASA's programs in Planetary Geology
\& Geophysics and in Origins of Solar systems.

%%%%%%%%%%%%%%%%%%%%%%%%%%%%%%%%%%%%%%%%%%%%%%%%%%%%%%%%%%%%%%%%%%%%%%%%%%
% Brouwer Integral Appendix      %
%%%%%%%%%%%%%%%%%%%%%%%%%%%%%%%%%%%%%%%%%%%%%%%%%%%%%%%%%%%%%%%%%%%%%%%%%
%\newpage
\appendix
\section{Appendix \ref{appendix_BI}}
\label{appendix_BI}

Consider the orbital evolution
of a particle trapped by a migrating planet's 
$j+k:j$ mean--motion resonance. The planet's disturbing function
contains the the resonant term 
$R_{res}=R_{res}(\alpha,e,\phi_{jk})$ where $\alpha$ is the 
planet/particle semimajor axis 
ratio for a particle having an eccentricity $e$ and a
resonance angle having the form
\begin{equation}
  \label{phi_jk}
  \phi_{jk}=(j+k)\lambda-j\lambda_p-k\tilde{\omega}
\end{equation}
where $\lambda$ and $\lambda_p$
are the particle and planet's mean longitude while $\tilde{\omega}$
is the particle's longitude of periapse \citep{MD99}.
The relevant Lagrange planetary equations are
\begin{mathletters}
  \label{planetary_eqns}
  \begin{eqnarray}
    \frac{da}{dt}&=&\frac{2}{na}\frac{\partial R}{\partial\epsilon}\\
    \frac{de}{dt}&=&-\frac{\sqrt{1-e^2}}{na^2e}\left[
      (1-\sqrt{1-e^2})\frac{\partial R}{\partial\epsilon}
      +\frac{\partial R}{\partial\tilde{\omega}}\right]
  \end{eqnarray}
\end{mathletters}
where $\epsilon=\lambda-nt$ is the particle's mean longitude at
epoch, $n$ is the particle's mean motion, $t$ is time, and
coplanar orbits are assumed.
According to the averaging principal, the non--resonant terms in the planet's
disturbing function are of high frequency and average out during the
particle's libration period, and the planet's disturbing function
is simply $R=R_{res}$. The derivatives of $R$ are
\begin{equation}
  \frac{\partial R}{\partial\epsilon}=
    (j+k)\frac{\partial R_{res}}{\partial\phi_{jk}}\quad\mbox{and}\quad
  \frac{\partial R}{\partial\tilde{\omega}}=
    -k\frac{\partial R_{res}}{\partial\phi_{jk}}.
\end{equation}
With these, the planetary equations (\ref{planetary_eqns}) can be combined into
a single differential equation,
\begin{equation}
  \frac{de}{dt}=-\frac{\sqrt{1-e^2}}{2e}\left(
    \frac{j}{j+k}-\sqrt{1-e^2}\right)\frac{1}{a}\frac{da}{dt}.
\end{equation}
It will be convenient to replace $e$ with the variable $E=\sqrt{1-e^2}$;
since $ede=-EdE$, the above equation can then be recast as
\begin{equation}
  \frac{da}{a}=\frac{2dE}{\gamma-E}
\end{equation}
where $\gamma=j/(j+k)$. This differential equation is now easily integrated
and yields $\ln a=-2\ln(\gamma-E)+\ln B$ where the integration
constant $B$ can also be expressed as
\begin{equation}
  \label{B}
  B=a\left(\sqrt{1-e^2}-\frac{j}{j+k}\right)^2
\end{equation}
({\it c.f.}, \citealt{HW95}). The earliest derivation of this integral of the 
motion known to us is given in \citet{B63}.

Brouwer originally used this integral to consider the motion 
of a particle being perturbed by a planet in a static orbit. However
\citet{YT99} recognized that this integral is preserved even
when the planet is migrating, which means that 
Eqn.\ (\ref{B}) can be used to predict the particle's 
eccentricity as its orbit is expanded by the planet's resonance:
\begin{equation}
  \label{e(a)}
  e^2(a)=1-\left(\frac{j}{j+k}+\sqrt{\frac{B}{a}}\right)^2
\end{equation}
If, prior to capture, the particle is in a circular orbit with
an initial semimajor axis $a_0$, then $B=a_0[k/(j+k)]^2$ and
\begin{equation}
  \label{e(a)2}
  e^2(a)=1-\left(\frac{j+k\sqrt{a_0/a}}{j+k}\right)^2,
\end{equation}
and the particle's eccentricity grows as the particle's orbit expands.
An approximate form of this expression is also derived in \citet{M93}.
Thus if Neptune's orbital expansion is indeed responsible for
the KBOs seen at Neptune's 3:2 resonance at $a=39.4$ AU
having eccentricities as high as $e\simeq0.32$ (see Fig.\ \ref{cold belt}), 
Eqn.\ (\ref{e(a)2}) indicates they must first have been orbiting
at $a_0=28.0$ AU at the time of capture, and that Neptune was initially 
orbiting at $a_N=a_0\gamma^{2/3}=21.4$ AU and hence migrated a distance 
$\Delta_N=8.7$ AU.

%%%%%%%%%%%%%%%%%%%%%%%%%%%%%%%%%%%%%%%%
% Ratios Appendix      %
%%%%%%%%%%%%%%%%%%%%%%%%%%%%%%%%%%%%%%%%
\section{Appendix \ref{appendix_ratios}}
\label{appendix_ratios}

The following will show that the ratio of the observed abundance of any two
dynamical classes of KBOs is approximately
equal to the ratio of the absolute abundances of the much larger unseen populations.
Begin by letting $N_x^{obs}(m)$ be the number of observed
KBOs that inhabit some dynamical class $x$ that are brighter than apparent 
magnitude $m$, where $x$ might represent, say, the 3:2 population. Also  
assume that all KBOs have the
same heliocentric distance $r$, which is a common assumption 
(e.g.,  \citealt{ITZ95, Bernstein04}) that simplifies this analysis 
considerably, and is  actually not a bad assumption since Fig.\ \ref{erosion} shows 
that most of the observed multi--opposition KBOs considered here do indeed 
inhabit a rather narrow belt, with more that half of these KBOs 
found at heliocentric distances within 3 AU of
$r=42$ AU. We will also assume that the Kuiper Belt has azimuthal symmetry,
which is justified further in Appendix \ref{symmetry}. We shall
also assume that all the KBO astronomers are observing this Belt largely
along the ecliptic, which also is not a bad assumption, since $72\%$
of the multi-opposition KBOs studied here have latitudes of $|\beta|<3^\circ$,
which is the typical inclination of the 
Belt's low--$i$ component \citep{B01}.
This indicates that the astronomers who discovered most of the KBOs in this
sample (represented by the red dots in Fig.\ \ref{hot belt})
were observing largely along the Belt's midplane. So, although these assumptions 
are not rigorously correct in detail, they are good
enough to allow us to assess qualitatively the relative abundances of the various
KBO populations. 

Let $\Sigma_x(m)$ represent the sky--plane number density of class--$x$ KBOs
({\it i.e.}, the cumulative luminosity function of class--$x$ KBOs), while
$\Sigma_x'(m)\equiv d\Sigma_x/dm$ be the {\em differential} luminosity function 
for class $x$. Then number of observed KBOs brighter than magnitude $m$ is
\citep{Bernstein04}
\begin{equation}
  N_x^{obs}(m)=\int\Sigma_x'(m)\Omega(m)\eta_x(m)dm,
\end{equation}
where $\Omega(m)$ is the total solid angle 
that has been surveyed by all KBO astronomers to a limiting magnitude 
$m$, and $\eta_x(m)$ is the efficiency at which a KBO of class $x$ and
magnitude $m$ and is detected. Thus the number of observed class $x$ KBOs of
having magnitudes in the interval $m\pm\Delta m/2$ is
\begin{equation}
  dN_x^{obs}(m)\equiv\frac{dN_x^{obs}}{dm}\Delta m=
    \Sigma_x'(m)\Omega(m)\eta_x(m)\Delta m.
\end{equation}
If we were to compare class $x$ KBOs to, 
say, class $y$ KBOs, then the ratio of their observed abundances is
\begin{equation}
  r_{x:y}(m)=\frac{dN_x^{obs}(m)}{dN_y^{obs}(m)}
    \simeq\frac{\Sigma_x'(m)}{\Sigma_y'(m)},
\end{equation}
where it is assumed that the KBO detection efficiency 
is insensitive to dynamical class, {\it i.e.}, $\eta_x(m)\simeq\eta_y(m)$.
Note that this ratio depends only on the KBOs' differential luminosity functions 
$\Sigma'(m)$, and that it is insensitive to their discovery details such as the total
solid angle $\Omega(m)$ that the KBO astronomers have surveyed to depth $m$.

The cumulative luminosity function for class $x$ can be written as
$\Sigma_x(m)=N_x(m)/\Omega_{total}$
where $\Omega_{total}$ is the Kuiper Belt's total solid angle, and
$N_x(m)$ is the cumulative magnitude distribution for class $x$, {\it ie,}
the total number of KBOs in class $x$ that are brighter than magnitude $m$. Thus
\begin{equation}
  \Sigma_x'(m)=\frac{1}{\Omega_{total}}\left(\frac{dN_x}{dm}\right)=
    \frac{1}{\Omega_{total}}\left(\frac{dN_x}{dR}\right)\left(\frac{dR}{dm}\right)
\end{equation}
since a body's magnitude $m$ is a function of its radius $R$ via Eqn.\ (\ref{mR}),
for which
\begin{equation}
  \frac{dN_x}{dR}=-\frac{QN_{x,total}}{R_{min}}
    \left(\frac{R}{R_{min}}\right)^{-(Q+1)}
\end{equation}
is the differential size distribution for class $x$
which has a total of  $N_{x,total}$ bodies having radii in the interval
$R_{min}<R<R_{max}$ (see Eqn.\ \ref{N(R)}). 
And since $dR/dm=-R/5$ (see Eqn.\ \ref{mR}),
the differential luminosity function can be written
\begin{equation}
   \Sigma_x'(m(R))=\frac{QN_{x,total}}{5\Omega_{total}}
    \left(\frac{R}{R_{min}}\right)^{-Q}.
\end{equation}
Thus if any two KBO classes $x$ and $y$ have the same size distribution $Q$,
then the ratio of their observed abundances is simply
\begin{equation}
  r_{x:y}(m)\simeq\frac{\Sigma_x'(m)}{\Sigma_y'(m)}\simeq
    \frac{N_{x,total}}{N_{y,total}}.
\end{equation}
In other words, the ratio of the observed abundance of any two classes of KBOs
is approximately equal to the ratio of their intrinsic abundances, provided all bodies
have the same size distribution $Q$. But if populations $x$ and $y$ have
different size distributions, then $r_{x:y}(m)$ will vary with $m$
(see Section \ref{census}).

%%%%%%%%%%%%%%%%%%%%%%%%%%%%%%%%%%%%%%%%
% Symmetry Appendix      %
%%%%%%%%%%%%%%%%%%%%%%%%%%%%%%%%%%%%%%%%
\section{Appendix \ref{symmetry}}
\label{symmetry}

\begin{figure}
\epsscale{1.0}
\plotone{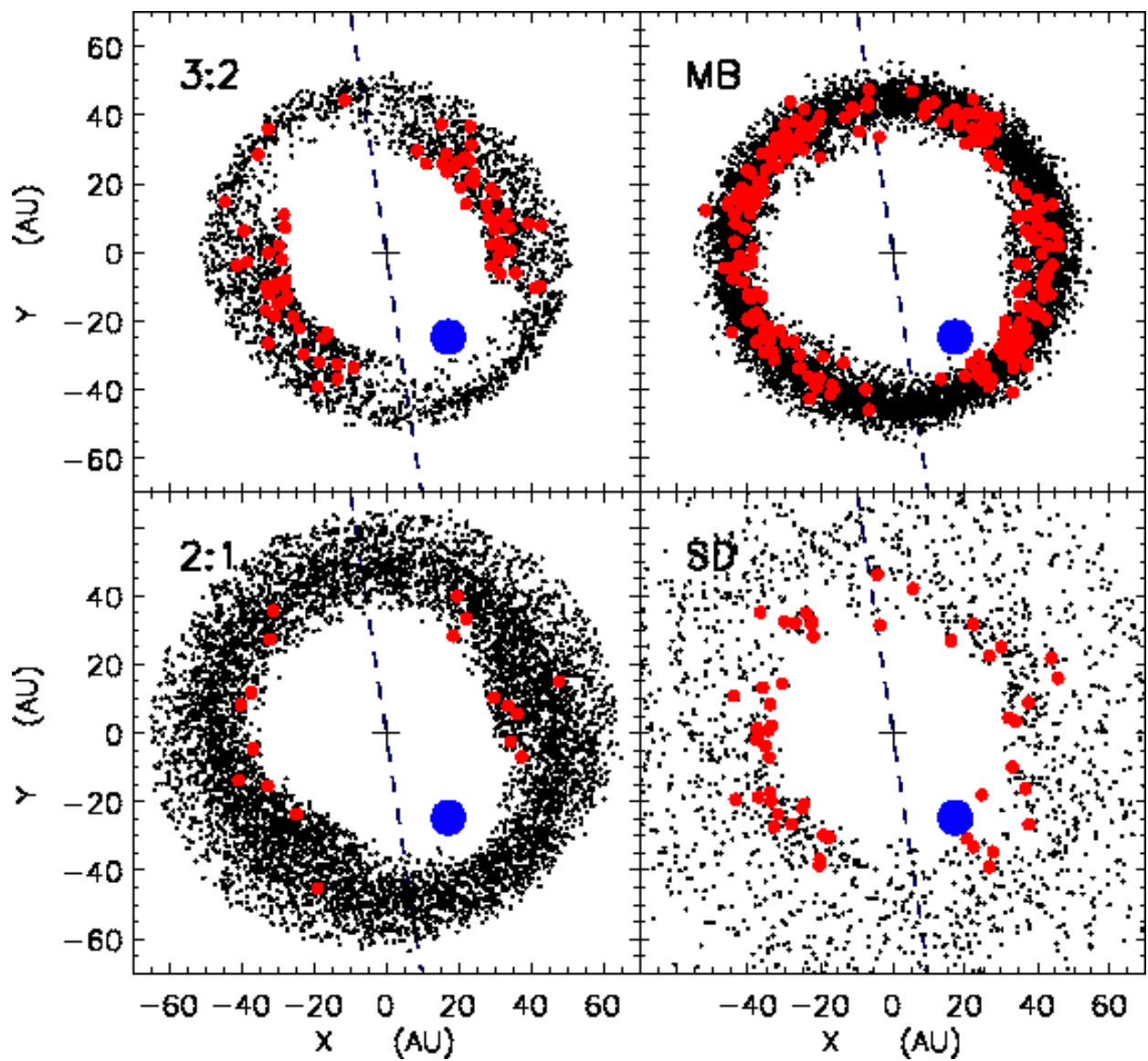}
\figcaption{
  \label{longitudes}
  Black dots indicate the positions of all simulated particles relative to Neptune
  recorded at all times later than $t=3$ Gyrs. Red dots indicate the ecliptic 
  coordinates of the observed multi-opposition KBOs, shown for July 1, 2000, 
  which is the date by which half of the sample considered here had been
  discovered. The blue dot is Neptune's position on this date, 
  and the dashed line shows where the 
  galactic plane penetrates the ecliptic $x$--$y$ plane. 
  The particles and KBOs are sorted by their
  dynamical membership: bodies in or very near the 3:2 resonance (upper left),
  Main Belt (MB, upper right), 2:1 (lower left), and the so--called Scattered Disk
  Objects (SD, lower right) having perihelia $30<q<40$ AU and $a>48.4$ AU. 
}
\end{figure}

The Monte Carlo model of Section \ref{census} replicates each
Nbody survivor seen in Fig.\ \ref{hot belt} $10^4$ times by randomizing
the particles' mean anomalies, which results in a model Kuiper Belt having 
azimuthal symmetry. Thus the model implicitly assumes that the
Kuiper Belt has this same symmetry. However
one might question this assumption,
since it is well known that a planet's resonant perturbations can rearrange the 
longitudes of a small body population. For instance,  particles at Neptune's 3:2
tend to approach perihelion at longitudes that are $\sim90^\circ$ away from the planet
\citep{M95}. This is illustrated in the upper left corner of Fig.\ \ref{longitudes}, 
which shows the ecliptic coordinates of all simulated particles that are in 
or very near Neptune's 3:2 resonance. This figure shows that Neptune
tends to arrange the 3:2 bodies preferentially away from the Sun--Neptune line,
causing regions that lead/trail Neptune by $\pm90^\circ$ to be more
densely populated, termed `sweet spots' in \cite{CJ02}. Figure  \ref{longitudes}
shows that resonant shepherding also occurs at Neptune's 2:1, again
similar to that seen in \cite{CJ02}. Red dots indicate the
positions of the multi--opposition KBOs considered
here, which shows that KBOs in or near the 3:2  resonance tend to be discovered
at longitudes that are roughly $\pm90^\circ$ away from Neptune, as expected. 

It is thus possible that Neptune's rearrangement of the Belt might
skew our estimate of the resonant populations. If, for example, KBO astronomers were
preferentially surveying the Belt along longitudes that are $\pm90^\circ$ away from 
Neptune, they would detect a rather high sky--plane number density of KBOs in the 3:2.
If we then assumed that this 3:2 number density were uniform about the entire ecliptic,
we would overestimate the total 3:2 population. Similarly, if astronomers systematically
observed the Belt towards Neptune or $180^\circ$ away, we would underestimate
the 3:2's average sky--plane density and undercount its total population. 
But if astronomers surveyed all longitudes with equal frequency, 
then these competing effects---due to 
Neptune pushing KBOs away from certain longitudes and towards other 
longitudes---should wash out, resulting in an 
estimate of the 3:2 population that is approximately reliable.

The upper right part of Fig.\ \ref{longitudes} shows the ecliptic coordinates of
the observed multi--opposition Main Belt KBOs. Since the Main Belt population is 
likely azimuthally symmetric, these red dots should
be a good indicator of where astronomers
are looking for KBOs. It is quite clear from this figure that these astronomers' 
lines--of--sight are not distributed uniformly about the ecliptic. For instance, these
KBO hunters tend to avoid the galactic plane, which passes through the ecliptic 
along the dashed line. The MB figure also shows that astronomers have avoided a 
narrow portion of the  3:2's `sweet spot' that leads Neptune by about $90^\circ$. But it
is also quite clear that these same astronomers are not systematically staring at the 3:2's 
sweet spots, nor are they systematically avoiding them.  Consequently, our simple visual
inspection of this KBO sample indicates that our estimate of the 3:2 population
will not be significantly baised towards under or overcounting the 3:2 population, 
and that the 3:2 abundance reported in Section \ref{calibration} is indeed representative. 
A similar conclusion is also drawn for the much sparser 2:1 population.

Of course, this treatment succeeds only if the Monte Carlo method of
Section \ref{census} does not alter the particles' radial distribution
as they have their mean anomalies randomized by the replication process.
However it is straighforward to show that this is indeed the case. For instance,
randomizing the mean anomalies of the 3:2 population seen Fig.\ \ref{longitudes} 
does not change their radial distribution in any significant way.

%%%%%%%%%%%%%%%%%%%%%%%%%%%%%%%%%%%%%%%%%%%%%%%%%%%%%%%%%%%%%%%%%%%%%%%%%
% References        %
%%%%%%%%%%%%%%%%%%%%%%%%%%%%%%%%%%%%%%%%%%%%%%%%%%%%%%%%%%%%%%%%%%%%%%%%%

%%%%%%%%%%%%%%%%%%%%%%%%%%%%%%%%%%%%%%%%%%%%%%%%%%%%%%%%%%%%%%%%%
% Table I.                                                     %
%%%%%%%%%%%%%%%%%%%%%%%%%%%%%%%%%%%%%%%%%%%%%%%%%%%%%%%%%%%%%%%%%
\newpage
\begin{deluxetable}{llll}
\renewcommand{\arraystretch}{0.7}
\tablenum{1}
\tablewidth{0pt}
\tablecaption{Abundance and mass of KBOs having radii $50<R<1000$ km
  assuming density $\rho=1$ gm/cm$^3$ and albedo $p=0.04$ \label{pop_table}}
\tablehead{
  \colhead{dynamical class\hspace*{15ex}} & \colhead{site (AU)\hspace*{15ex}} & 
  \colhead{population} & \colhead{mass (M$_\oplus$)}
}
\startdata
Centaurs   & $a<a_N$  & $<80$   & $<4\times10^{-5}$ \\
Trojans    & $a=a_N$  & $<1.1\times10^3$ & $<5\times10^{-4}$ \\
3:2    & $a=39.5$  & $2.7\times10^3$ & $3.1\times10^{-3}$ \\
Main Belt    & $40.1\le a\le 47.2$ & $1.3\times10^5$ & 0.059   \\ 
2:1    & $a=47.8$  & $5.3\times10^3$ & $2.4\times10^{-3}$ \\
Scattered Disk   & $50<a<150\ \&\ 28<q<40$ & $2.5\times10^4$ & 0.011 
\enddata
\end{deluxetable}

%%%%%%%%%%%%%%%%%%%%%%%%%%%%%%%%%%%%%%%%%%%%%%%%%%%%%%%%%%%%%%%%%
% Figures & captions.                                           %
%%%%%%%%%%%%%%%%%%%%%%%%%%%%%%%%%%%%%%%%%%%%%%%%%%%%%%%%%%%%%%%%%
\newpage

%Figure 1
\begin{figure}
\epsscale{1.0}
\plotone{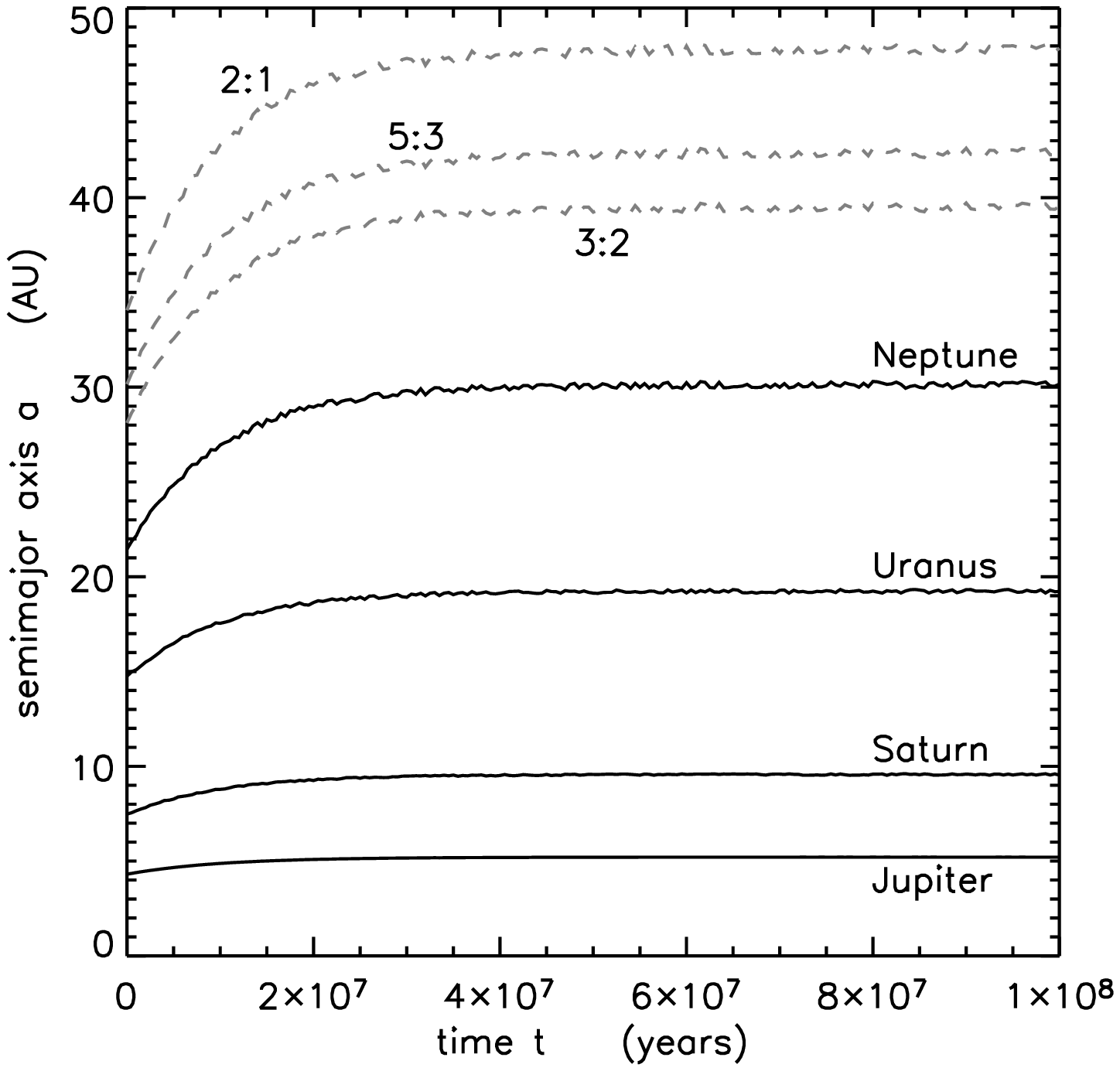}
\end{figure}

%Figure 1 caption
\begin{figure}
\figcaption{
\label{migrate}
The planets' semimajor axes $a_j$ versus time t, as well as
a few of Neptune's mean  motion resonances that are effective at 
trapping particles.}
\end{figure}

%Figure 2
\begin{figure}
\epsscale{1.0}
\plotone{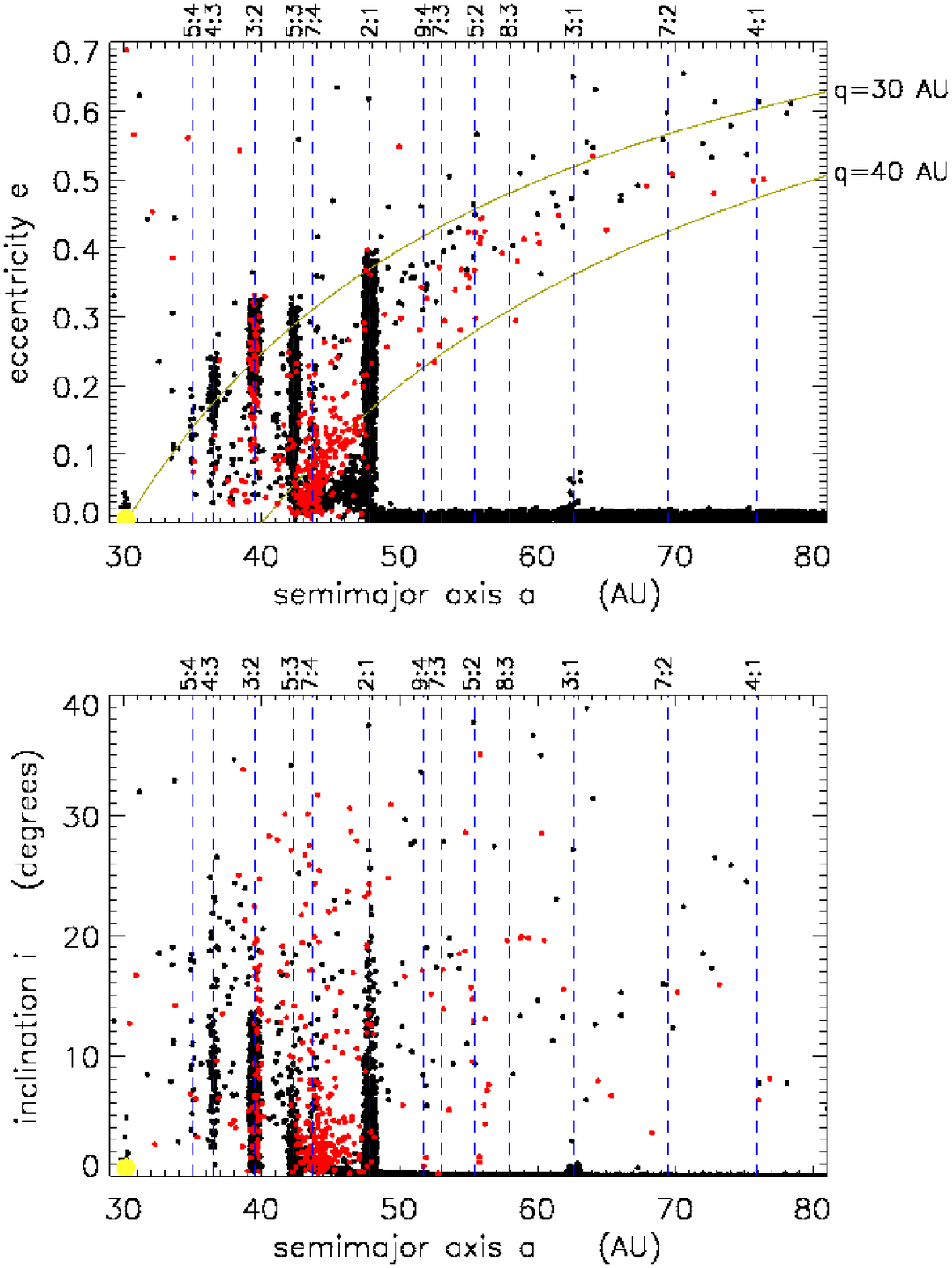}
\end{figure}

%Figure 2 caption
\begin{figure}
\figcaption{
\label{cold belt}
A simulation of Neptune's migration into a dynamically cold Kuiper Belt.
The model Kuiper Belt is initially composed of $10^4$ massless particles
having semimajor axes randomly distributed over $20<a<80$ AU
with a surface number density that varies as $a^{-2}$.
The particles initial eccentricities are Rayleigh distributed about a
mean $\langle e\rangle=0.001$,
and their initial inclinations (which are measured with respect to the
system's invariable plane) are similarly distributed over 
$\langle \sin i\rangle=\langle e\rangle/2$, while
their other angular orbit elements are uniformly distributed over $2\pi$.
The four giant planets' orbits migrate according to the prescription 
described in Section \ref{model}, and the black dots shows the simulated
Kuiper Belt endstate---the particles' $e$'s and $i$'s versus $a$ 
at time $t=5\times10^8$ years. The red dots are the ecliptic orbit elements
of KBOs reported by the Minor Planet Center
as having been observed for at least two oppositions.
The yellow dots indicate Neptune's orbit, and the vertical dashes show
the locations of Neptune's various $j+k:j$ mean--motion resonances. 
Orbits having perihelia $q=30$ and 40 AU are also indicate by the curves.}
\end{figure}

%Figure 3
\begin{figure}
\epsscale{1.0}
\vspace*{-5ex}\plotone{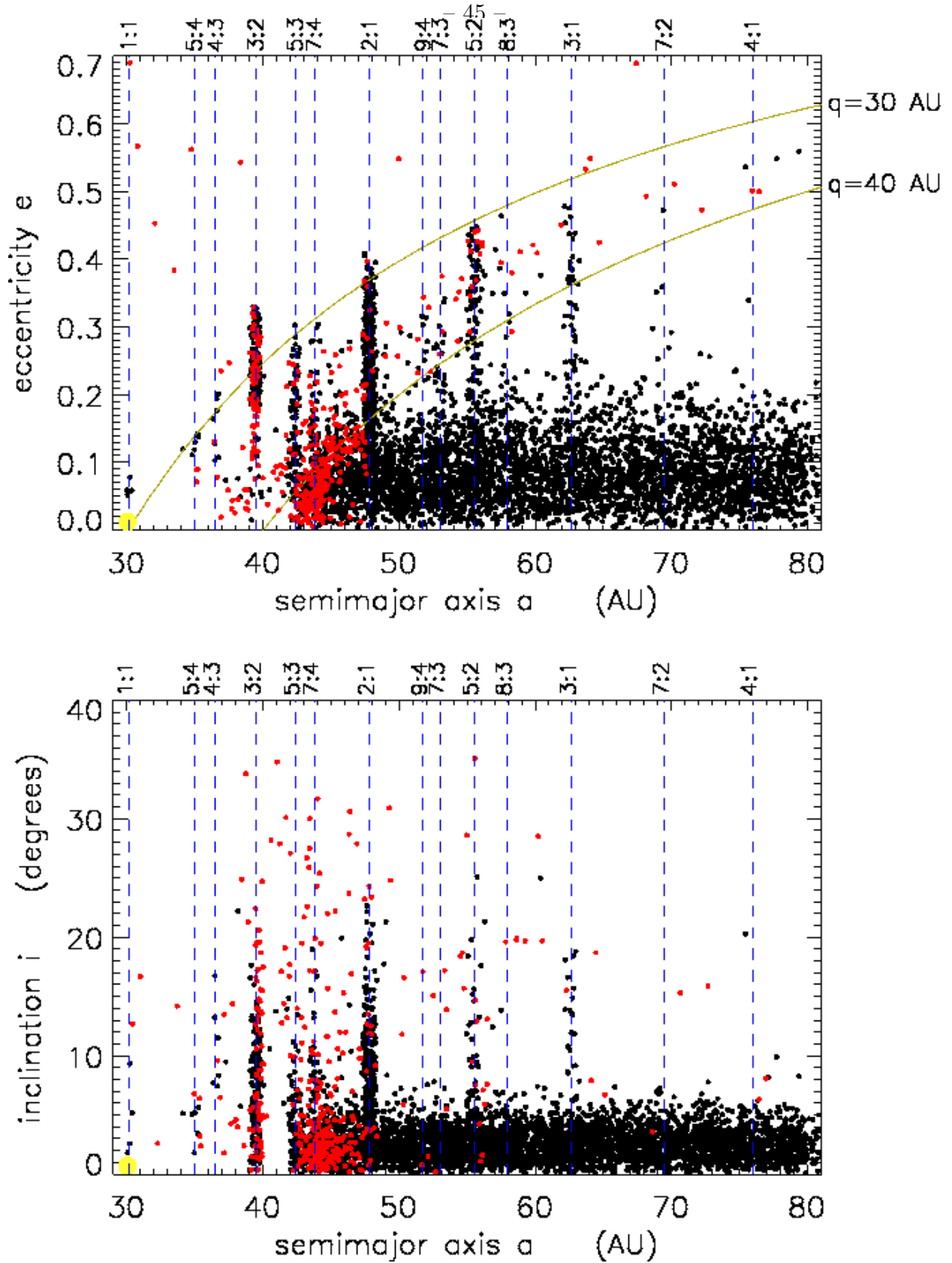}
\vspace*{-5ex}\figcaption{
  \label{hot belt}
  Neptune migrates into a dynamically hot disk of $N=10^4$ particles having
  initial $e$'s Rayleigh distributed about $\langle e\rangle=0.1$ and 
  inclinations similarly distributed about 
  $\langle \sin i\rangle=\langle e\rangle/2$,
  while all other initial orbits are distributed as in Fig.\ \ref{cold belt}.
  This system is evolved for $t=4.5\times10^9$ years.
}
\end{figure}

%Figure 4
\begin{figure}
\epsscale{1.0}
\plotone{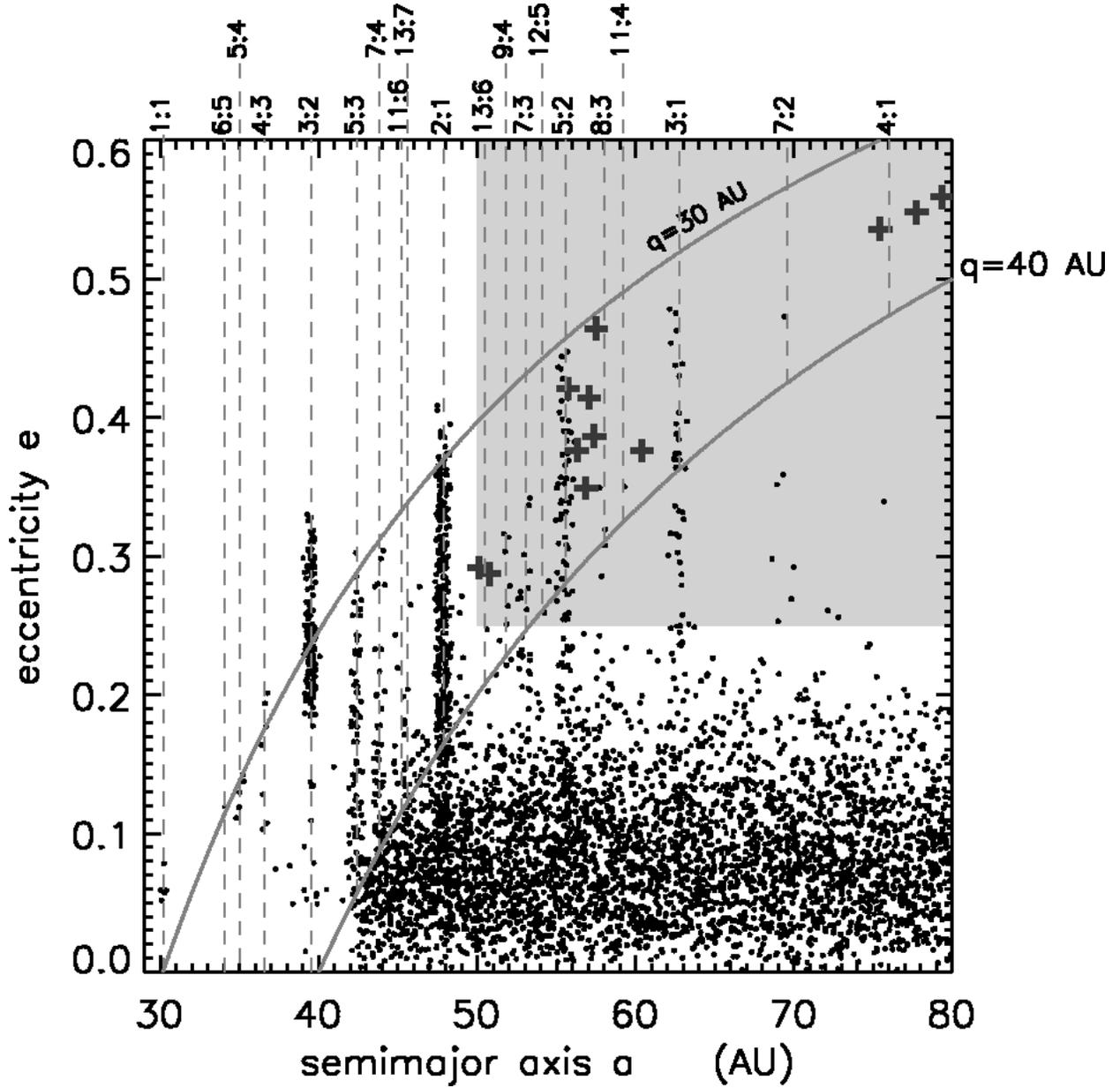}
\figcaption{
  \label{sd}
  Eccentricity $e$ versus semimajor axis $a$ at time $t=4.5\times10^9$ years
  for the model population having initial $e\sim0.1$ shown in Fig.\ \ref{hot belt}.
  The vertical dashes indicate
  the mean motion resonances that are occupied by trapped particles with
  perihelia $q<40$ AU or particles inhabiting resonances in
  the gray region with libration amplitudes $|\phi_{jk}|\le90^\circ$. 
  Crosses indicate scattered particles.
}
\end{figure}

%Figure 5
\begin{figure}
\epsscale{1.0}
\plotone{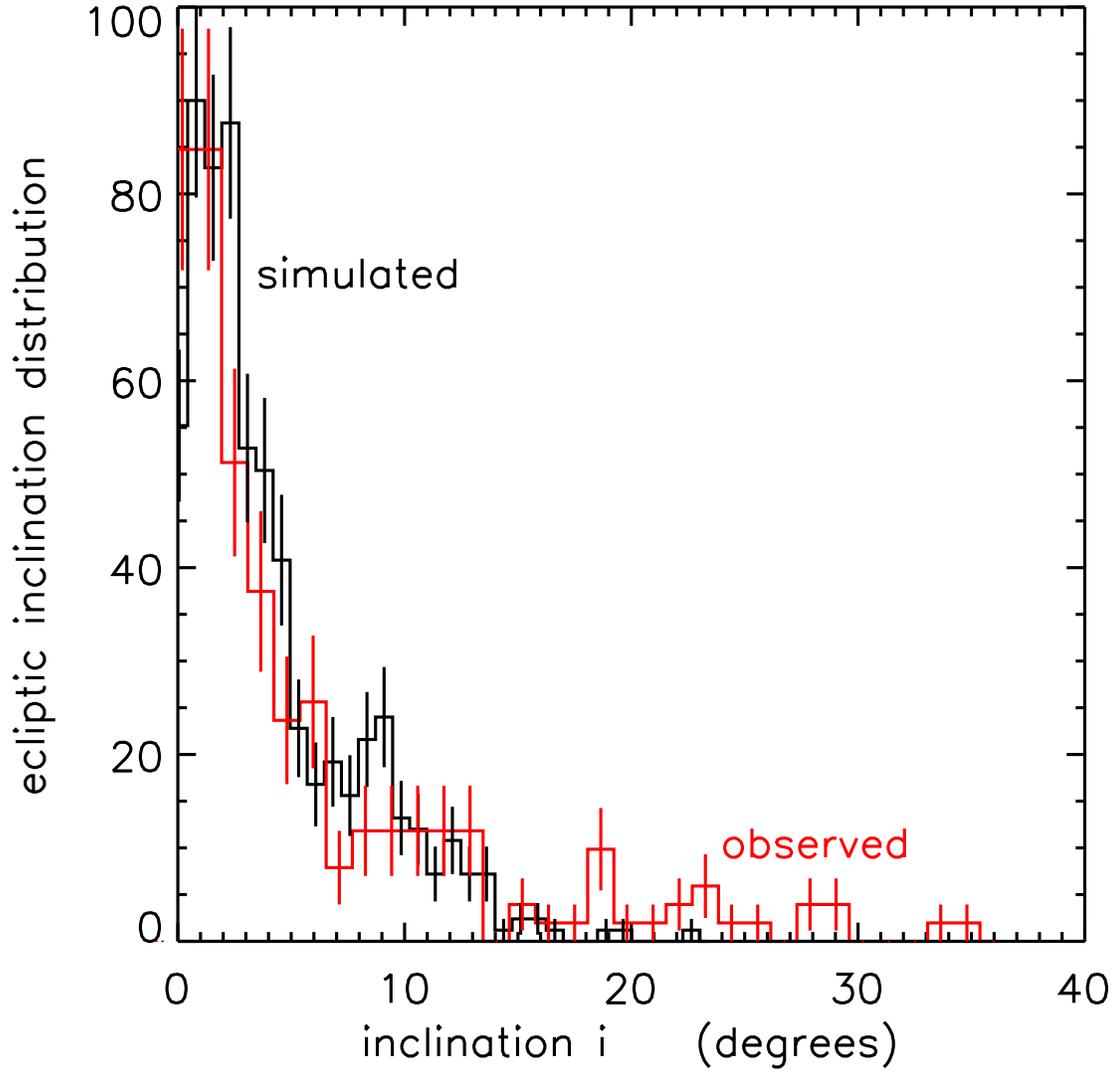}
\figcaption{
  \label{idist}
  The ecliptic inclination distributions for the simulated and observed bodies
  in Fig.\ \ref{hot belt} that have perihelia $q<42$ AU and latitudes 
  $|\beta|\le1^\circ$. Error bars give the Poisson counting uncertainty,
  and the vertical scale is arbitrary.
}
\end{figure}

%Figure 6
\begin{figure}
\epsscale{1.0}
\plotone{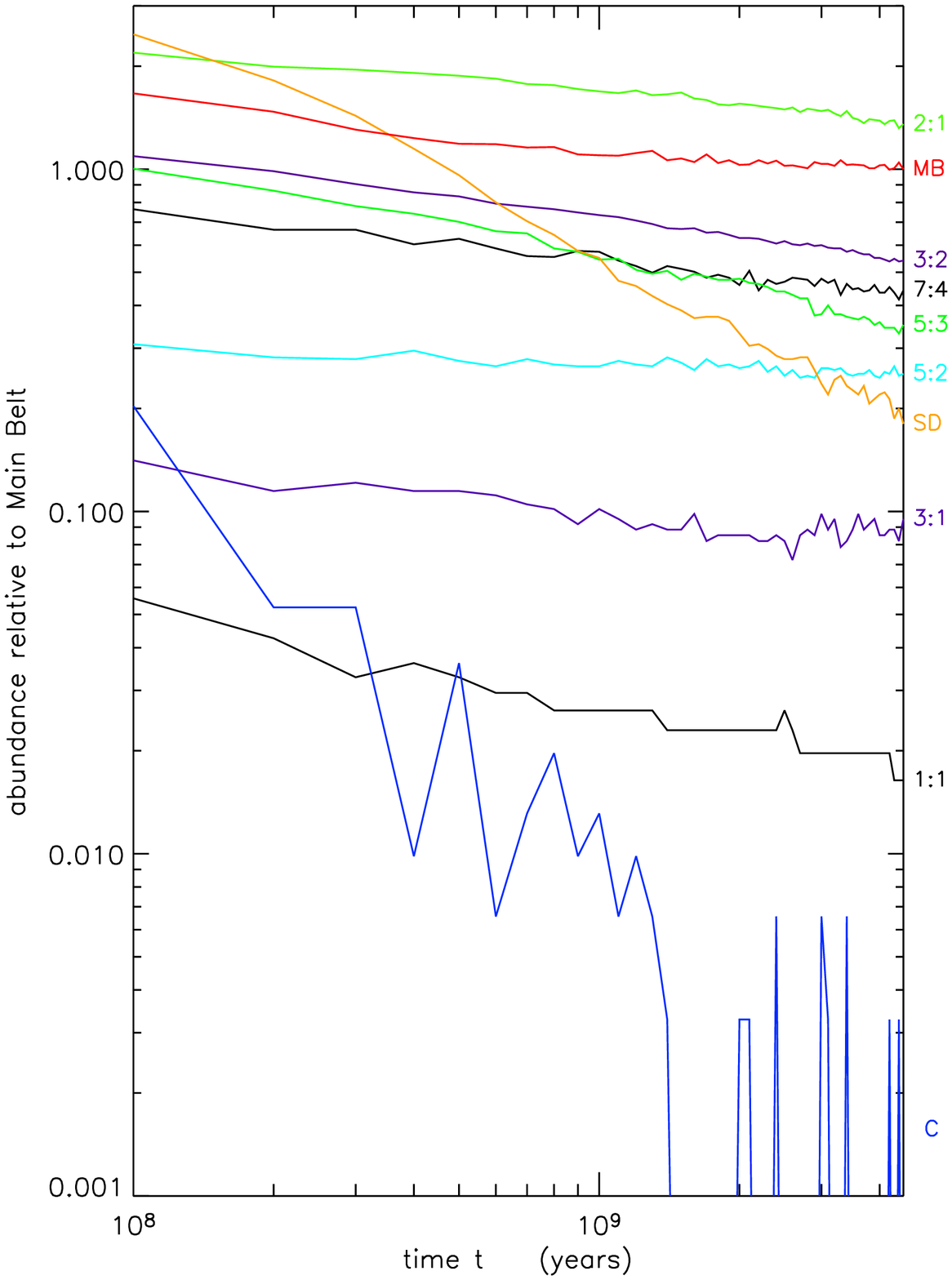}
\end{figure}

%Figure 6 caption
\begin{figure}
\figcaption{
  \label{abundance}
  The abundances of the Belt's various dynamical populations are plotted over time 
  $t$, with all populations normalized to the Main Belt (MB) population
  at time $t=4.5\times10^9$ years, and counting only those particles having perihelia
  $q\le45$ AU. The resonant populations are counts of all particles having
  semimajor axes $\pm0.6$ AU within exact resonance, while the MB curve gives the 
  number of particles having semimajor axes $40.1\le a\le 47.2$ AU, excepting
  bodies in/near the 5:3 and 7:4 resonances. However the Scattered Disk
  (SD) curve is the number of particles having $a>49$ AU and perihelia $q<40$ AU
  that are not members of any of the indicated resonances (e.g., the 5:2 and 3:1),
  while Centaurs (C) are particles having $a<$ Neptune's semimajor axis.
}
\end{figure}

%Figure 7
\begin{figure}
\epsscale{1.0}
\plotone{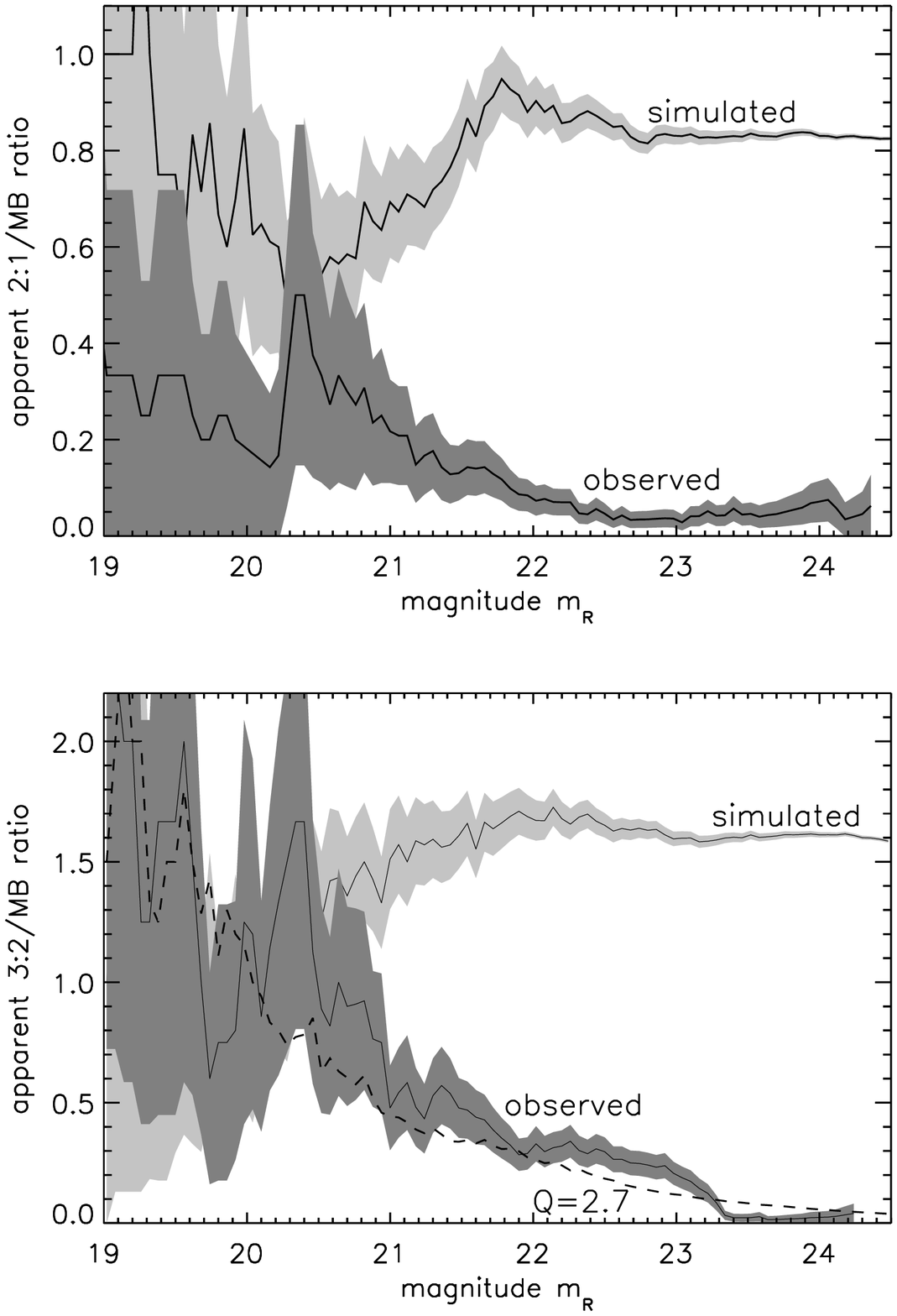}
\end{figure}

%Figure 7 caption
\begin{figure}
\figcaption{
  \label{ratios}
  The dark--gray curve in the upper plot is the number of KBOs that have been 
  observed (for two or more oppositions) orbiting
  in/near the 2:1 resonance having apparent magnitudes $m_R\pm\Delta m$
  relative to the observed Main Belt (MB) population at $40.1\le a\le 47.2$ AU,
  while the lower dark--gray curve is the observed 3:2/MB ratio.
  Likewise, the light--gray curves are the 2:1/MB and 3:2/MB
  ratios for the simulation of Fig.\ \ref{hot belt}, where these bodies' sizes and 
  magnitudes are assigned using the Monte Carlo method of Section \ref{census}
  assuming $Q=4.4$ and $R_{min}=20$ km. All of these curves are smoothed over a 
  magnitude--window having a half--width $\Delta m=0.5$, and the vertical half--width
  of the gray zones are one standard deviation assuming Poisson counting
  uncertainties. The observed curves end at $m_R\simeq24.5$, which is the
  magnitude of the faintest multi--opposition KBO. The dashed curve is 
  the simulated 3:2/MB ratio when the 3:2 population has
  a shallow size distribution with $Q=2.7$ and $R_{min}=4.3$ km, while the 
  MB bodies have $Q=4.4$ and $R_{min}=20$ km.
}
\end{figure}

%Figure 8

%Figure 9

%Figure 10
\clearpage

%Figure 11

%Figure 12
\begin{figure}
\epsscale{1.0}
\plotone{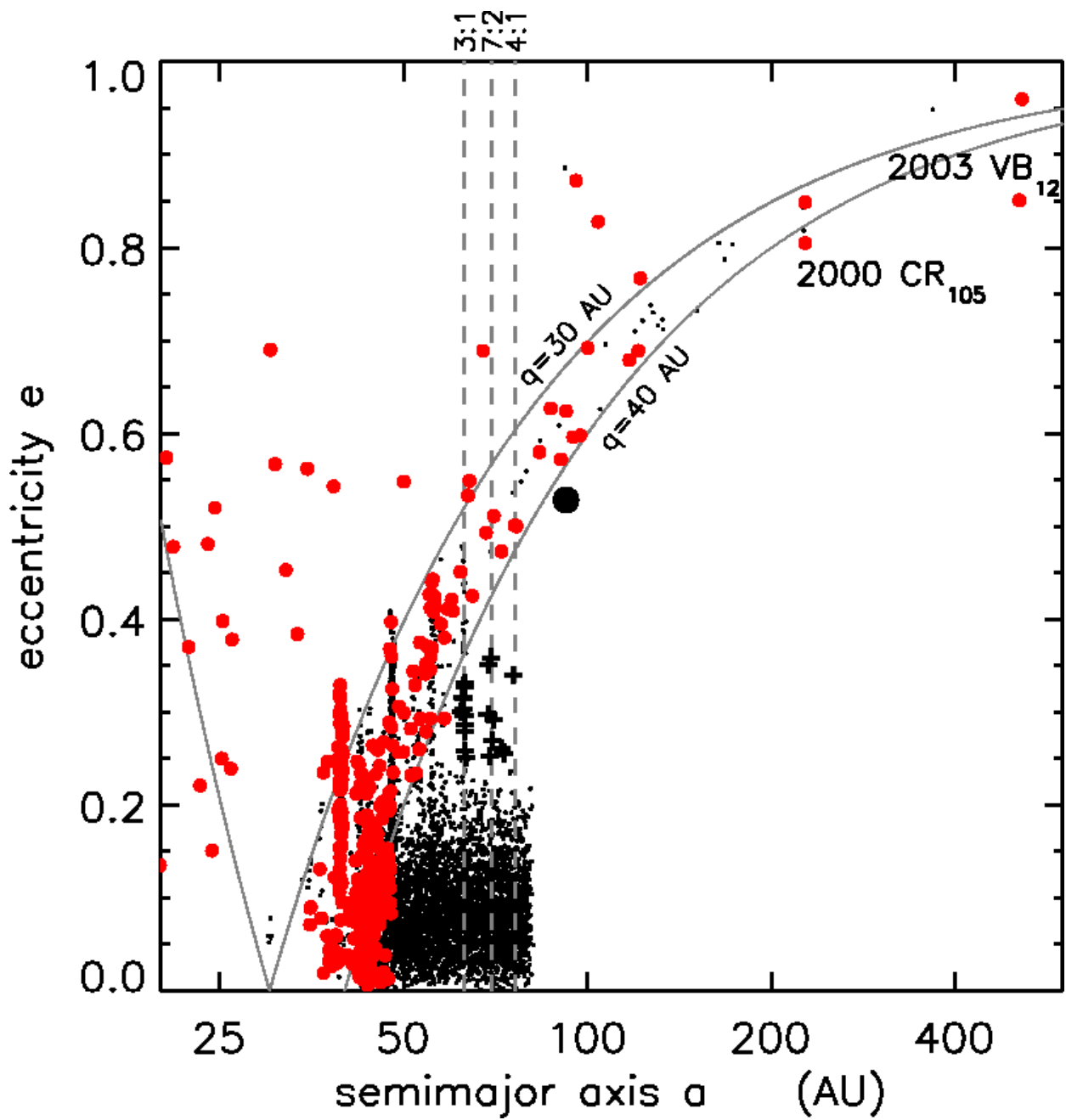}
\figcaption{
  \label{oss}
  Eccentricities $e$ are plotted versus $a$ on a logarithmic
  axis for the particles of Fig.\ \ref{hot belt}.
  Red dots are the KBOs observed over multiple oppositions,
  and the curves indicate the $q=40$ AU threshold as well as Neptune--crossing
  orbits. Two members of the so--called extended Scattered Disk,
  2000 CR$_{105}$ and 2003 VB$_{12}$, are also indicated. Crosses indicate
  particles resonantly trapped at the 3:1, 7:2, and the 4:1. The large black dot 
  indicates the only simulated particle that was scattered into a high--perihelia
  orbit ($q=43.5$ AU) that resembles 2000 CR$_{105}$.
}
\end{figure}

%Figure 13
\begin{figure}
\epsscale{0.75}
\plotone{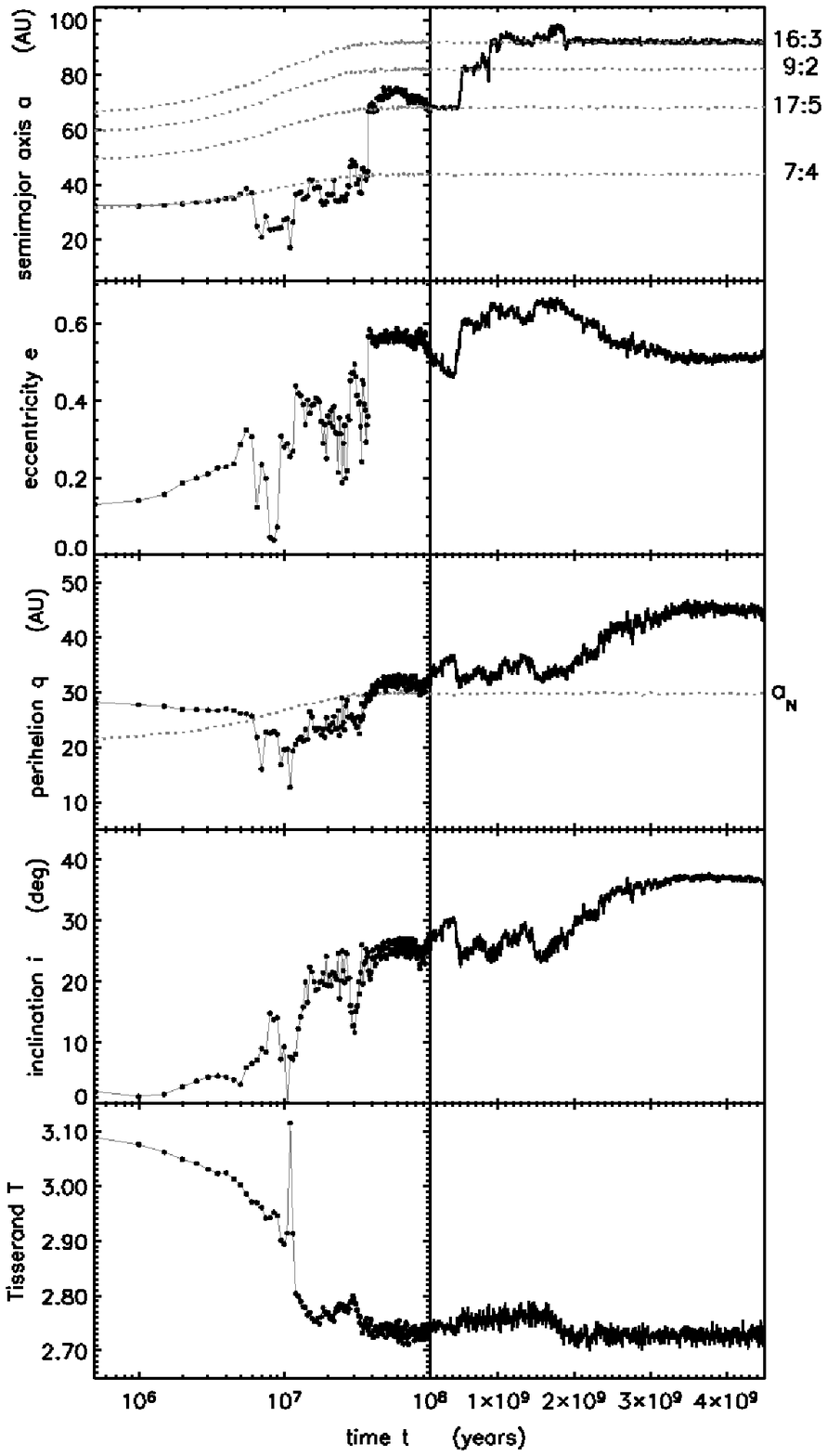}
\end{figure}

%Figure 13 caption
\newpage
\begin{figure}
\figcaption{
  \label{p2847}
  The orbital history of the only simulated scattered object that inhabits
  the extended Scattered Disk (e.g., the large black dot in Fig.\ \ref{oss}).
  The particle's $a$, $e$, $q$ , $i$, and Tisserand parameter
  $T=a_N/a+2\sqrt{(a/a_N)(1-e^2)}\cos i$ are first plotted versus logarithmic
  time (left half of the Figure) and then linearly (right half). The dotted
  curves are Neptune's semimajor axis $a_N$ and a few of its mean--motion resonances.
}
\end{figure}

%Figure 14
\newpage

%Figure 15

%Figure 16

\end{document}